**Astronomy & Astrophysics**

# Peering into the young planetary system AB Pic

## Atmosphere, orbit, obliquity, and second planetary candidate

P. Palma-Bifani[1,2,3], G. Chauvin[2,3,4], M. Bonnefoy[4], P. M. Rojo[1], S. Petrus[4,5], L. Rodet[6], M. Langlois[7], F. Allard[7], B. Charnay[8], C. Desgrange[4,11], D. Homeier[15,16], A.-M. Lagrange[8], J.-L. Beuzit[9], P. Baudoz[8], A. Boccaletti[8], A. Chomez[8,4], P. Delorme[4], S. Desidera[10], M. Feldt[11], C. Ginski[12], R. Gratton[10], A.-L. Maire[4], M. Meyer[13,14], M. Samland[10], I. Snellen[12], A. Vigan[9], and Y. Zhang[12]

[1] Unidad Mixta Internacional Franco-Chilena de Astronomía, CNRS/INSU UMI 3386, and Departamento de Astronomía,
   Universidad de Chile, Casilla 36-D, Santiago, Chile
[2] Departamento de Astronomía, Universidad de Chile, Casilla 36-D, Santiago, Chile
[3] Laboratoire Lagrange, Université Côte d'Azur, CNRS, Observatoire de la Côte d'Azur, 06304 Nice, France
   e-mail: paulina.palma-bifani@oca.eu
[4] Univ. Grenoble Alpes, CNRS, IPAG, 38000 Grenoble, France
[5] Núcleo Milenio Formación Planetaria – NPF, Universidad de Valparaíso, Av. Gran Bretaña 1111, Valparaíso, Chile
[6] Cornell Center for Astrophysics and Planetary Science, Department of Astronomy, Cornell University, Ithaca, NY 14853, USA
[7] Univ Lyon, ENS de Lyon, Univ Lyon 1, CNRS, Centre de Recherche Astrophysique de Lyon UMR5574, 69007 Lyon, France
[8] LESIA, Observatoire de Paris, PSL Research University, CNRS, Sorbonne Université, UPMC Univ. Paris 06, Univ. Paris Diderot,
   Sorbonne Paris Cité, 5 place Jules Janssen, 92195 Meudon, France
[9] Aix-Marseille Univ., CNRS, CNES, LAM, 38 rue Frédéric Joliot-Curie, 13388 Marseille Cedex 13, France
[10] INAF – Osservatorio Astronomico di Padova, Vicolo dell' Osservatorio 5, 35122 Padova, Italy
[11] Max-Planck-Institut für Astronomie, Königstuhl 17, 69117 Heidelberg, Germany
[12] Leiden Observatory, Leiden University, 2300 RA Leiden, The Netherlands
[13] Institute for Particle Physics and Astrophysics, ETH Zürich, Wolfgang-Pauli-Strasse 27, 8093 Zürich, Switzerland
[14] Department of Astronomy, University of Michigan, 1085 S. University Ave, Ann Arbor, MI 48109, USA
[15] Aperio Software Ltd., Insight House, Riverside Business Park, Stoney Common Road, Stansted, Essex, CM24 8PL, UK
[16] Förderkreis Planetarium Göttingen, Göttingen, Germany

**ABSTRACT**

*Aims.* We aim to revisit the formation pathway of AB Pic b, an imaged companion that straddles the exoplanet/brown-dwarf boundary. We based this study on a rich set of observations, which allows us to investigate its orbital and atmospheric properties.
*Methods.* We composed a spectrum of AB Pic b by merging archival medium-resolution (~4000) VLT/SINFONI K band (1.96–2.45 μm) data with published spectra at *J* and *H* bands from Magellan-AO/CLIO2, *Lp* band from SINFONI, and photometric measurements from HST (visible) and *Spitzer* (mid-infrared). We modeled the spectrum with ForMoSA, following a forward-modeling approach based on two atmospheric models: ExoREM and BT-SETTL13. In parallel, we determined the orbital properties of AB Pic b fitting orbital solutions to astrometric measurements from NaCo (2003 and 2004) and SPHERE (2015).
*Results.* The orbital solutions favor a semi-major axis of $190^{+200}_{-50}$ au on a highly inclined orbit (edge-on), but with a poorly constrained eccentricity. From the atmospheric modeling with Exo-REM, we derive an effective temperature of $1700 \pm 50$ K and surface gravity of $4.5 \pm 0.3$ dex, which are consistent with previous findings, and we report for the first time a $^C/_O$ ratio of $0.58 \pm 0.08$, consistent with the value for the Sun. The posteriors are sensitive to the wavelength interval and the family of models used. Given the published rotation period of 2.1 h and our derived $v\sin(i)$ of $73^{+11}_{-27}$ km s$^{-1}$, we estimate for the first time the true obliquity of AB Pic b to be between 45 and 135 deg, indicating a rather significant misalignment between the spin and orbit orientations of the planet. Finally, the existence of a proper-motion anomaly between the HIPPARCOS and *Gaia* Early Data Release 3 compared to our SPHERE detection limits and adapted radial velocity limits indicates the potential existence of a ~$6\,M_{Jup}$ inner planet orbiting from 2 to 10 au (40–200 mas).
*Conclusions.* The possible existence of an inner companion and the likely misalignment of the spin-axis orientation strongly favor a formation path by gravitational instability or core accretion within a protoplanetary disk at a smaller orbital radius followed by a dynamical interaction which scattered AB Pic b to its current location. Confirmation and characterization of this unseen inner exoplanet and access to a broader wavelength coverage and higher spectral resolution for the characterization of AB Pic b will be essential for probing the uncertainties associated with the atmospheric and orbital parameters.

## 1. Introduction

Transit, direct imaging, and cross-dispersed spectroscopy have successfully allowed us to describe the atmospheres of exoplanets. Since the first characterization of HD 209458 b, an exoplanet classified as a hot Jupiter (Charbonneau 2002), a broad diversity of physical processes have been found to be at play. Direct imaging studies have mainly focused on young, self-luminous giant gaseous planets, which are not strongly irradiated and orbit their host star at relatively large separations (>100 mas, >5–10 au for a typical young star at 50–100 pc), allowing high levels of complementarity between transit and cross-dispersed studies of Hot Jupiters and the classical spectroscopic characterization of young isolated planemos or free-floating exoplanets (Jameson et al. 2008; Allers & Liu 2013). Emblematic systems such as HR 8799 (Marois et al. 2010), β Pictoris (Lagrange et al. 2019), HD 95086 (Rameau et al. 2013), 51 Eri (Macintosh et al. 2015), HIP 65426 (Chauvin et al. 2017), and PDS 70 (Keppler et al. 2018) offer a rich opportunity to investigate the architectures of young Solar System analogs, but also the atmospheres of young super-Jupiters sometimes still accreting material (Haffert et al. 2019). The architectures of these emblematic systems with directly imaged planets are observable in Fig. 1, together with the star and planet masses and spectral types.

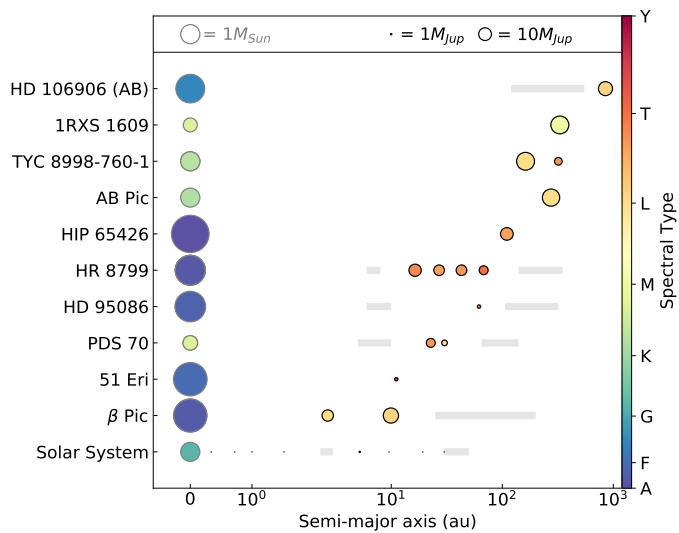

**Fig. 1.** Overview of the architectures of emblematic systems with imaged exoplanets. The sizes of the symbols coincide with masses of the objects. The sizes of the planet markers have been enhanced 70 times with respect to those of the star markers. Colors represent the spectral types, scaled by a power law to visualize details on low-mass objects. The gray area represents the disk or asteroid belt's location. The Solar System architecture is in the bottom row, where only Jupiter is scaled in mass consistently with the exoplanets.

Today's planet imagers, such as SPHERE at the Very Large Telescope (VLT; Beuzit et al. 2019), GPI at Gemini (Macintosh et al. 2006, 2015), and SCExAO/CHARIS at Subaru (Groff et al. 2015; Jovanovic et al. 2015), have yielded low-resolution ($R_\lambda$ = 30–90) emission spectra and photometry for tens of exoplanets exhibiting broad unresolved molecular bands ($H_2O$, $CH_4$, VO, FeH, CO) in the near-infrared (NIR: 0.95–2.45 µm). These data can be compared to predictions of atmospheric models to infer first-order information on their physical properties; mainly effective temperature ($T_{eff}$), surface gravity ($\log(g)$), or luminosity ($\log(L/L_\odot)$). However, they fail to provide crucial information at medium or high resolution ($R_\lambda > 1000$) that could enable a detailed exploration of fundamental parameters such as metallicity ([M/H]), carbon-to-oxygen ratio ($^C/_O$), and carbon isotopologic ratios ($^{12}C/^{13}C$). Adaptive-optics(AO)-fed integral-field spectrographs (IFS) operating at a medium spectral resolving power of $R_\lambda$ = 2000–5000 in the NIR, such as SINFONI at the VLT, NIFS at Gemini-North, and OSIRIS at Keck, already allow us to further our understanding of atmospheres. For example, those observations allow us to get information on the radial velocity (RV) of the studied objects (Ruffio et al. 2019, 2021) and are essential for preparing atmospheric models and theories for the upcoming generation of telescopes. With high resolution ($R_\lambda > 10\,000$), exploration of the RV and rotational velocity of exoplanets, which are connected to their spin and three-dimensional orbital properties, becomes feasible (Snellen et al. 2014). So far, a few directly imaged exoplanets have been characterized at this high spectral resolution, such as β Pictoris b (Snellen et al. 2014), GQ Lupi b (Schwarz et al. 2016), HR 8799 planets (Wang et al. 2021b), and HD 106906 b (Bryan et al. 2021). Young giant exoplanets at wide orbits offer unique benchmark laboratories with which to explore the spectral diversity connected to formation and evolution scenarios by investigating their atmospheres, physical properties, and system dynamics and stability in connection with the properties of the stellar host (Nowak et al. 2020; Molliere & Snellen 2019). Therefore, the previously described targets are and will be prime targets for upcoming telescopes, such as the *James Webb* Space Telescope (JWST[1], GTO target) and the Extremely Large Telescope (ELT[2], first light 2027).

Regarding the prime targets, differentiating substellar objects known as brown dwarfs (BD) from massive Jupiter-like exoplanets is difficult, because these populations overlap in observable properties. On the one hand, if a substellar object ($< 75\,M_{Jup}$) is found isolated in space, it is identified as a BD (Nayakshin 2017), and its atmospheric chemical composition may not differ from the interstellar medium (ISM; Chabrier et al. 2014). On the other hand, if the object is orbiting another star, it will be identified as a companion, but at present it is impossible to be certain about its formation history. The most accepted formation theories are gravoturbulent fragmentation of molecular clouds (Padoan & Nordlund 2004), core accretion (CA; Pollack et al. 1996), and gravitational instability (GI; Boss 1997). The observable properties of exoplanets and BD potentially relate to their formation environments, and mainly to the location of formation and composition of the accreted material that leaves an imprint on the upper atmospheric layers (Madhusudhan 2019).

By studying the atmospheres of young exoplanets and substellar objects, we expect to unveil their history and nature, which can be done by modeling their spectra. Atmospheric modeling can be achieved using two techniques: retrieval methods (Lavie et al. 2017) and forward modeling (Petrus et al. 2020, 2021). For this project, we used the latter where pre-computed self-consistent atmospheric models are compared to observational spectra following stochastic methods. The models are parameterized by the key atmospheric parameters and are received in the form of a grid. The main advantages of this approach are the significantly shorter computational time and the fact that the models provide a physical description of the processes at play. Different atmospheric grids are currently available, such as BT-SETTL13 (Allard et al. 2013), Exo-REM (Charnay et al. 2021), ATMO 2021 (Phillips et al. 2020), and Drift PHOENIX (Helling et al. 2008). Regarding the models, attempts to include clouds, hazes, winds, disequilibrium chemistry, and thermal inversions, among others, have been made in order to reproduce the rich set of emission and absorption lines. Today, many models are succeeding in reproducing the atmospheric spectral features, but given the complexity of these systems, attempts to develop 3D models and other improvements are also being made, which will likely lead to a revolution of this topic in the coming years (Fortney et al. 2021).

The fundamental parameters of the atmospheres studied in this work are $T_{eff}$, $\log(g)$, [M/H], and $^C/_O$. This latter particularly attracts our attention, because it has been shown to be constant on the ISM, varying mainly due to galactic chemical evolution (Chiappini et al. 2003), but to vary radially for gas and dust in protoplanetary disks triggered by the snow lines of $H_2O$, $CO_2$, and CO (Öberg et al. 2011). In a simple picture, massive planets formed by CA accrete their gaseous envelopes at the final formation stages, impacting the carbon and oxygen abundances in their atmospheres (Madhusudhan 2019). However, a companion formed by GI or gravoturbulent fragmentation is expected to have a $^C/_O$ similar to that of the ISM. Therefore, $^C/_O$ is proposed as a formation tracer, but a complete analysis must consider migration mechanisms, among others, to unveil the history (see Mollière et al. 2022 for a detailed discussion). An additional tracer recently proposed is the carbon monoxide isotopolog ratio (Molliere & Snellen 2019). Zhang & Snellen (2021) measured a value similar to that of the ISM ($^{12}CO/^{13}CO = 97^{+25}_{-18}$) for the atmosphere of a young isolated BD, while Zhang et al. (2021) studied TYC 8998-760-1, a wide-orbit (150 au) companion, and reported $^{12}CO/^{13}CO = 31^{+17}_{-10}$, explained by significant accretion of $^{13}C$-enriched ice beyond the CO snow line.

Here, we aim to investigate AB Pic b. This target is a substellar companion to AB Pic, a young, ~solar-type (K1V) star located at $50.14 \pm 0.04$ pc from the Sun based on *Gaia* eDR3 (Gaia Collaboration 2021). The star was recently re-classified as part of the younger Carina association by Booth et al. (2021) based on *Gaia* DR2 data where an age of $13.3^{+1.1}_{-0.6}$ Myr was estimated. This wide-orbit exoplanet (projected separation of $273 \pm 2$ au) was revealed by high-contrast coronographic images obtained with NaCo at VLT by Chauvin et al. (2005) and currently, the reported mass is $10 \pm 1\ M_{Jup}$. Bonnefoy et al. (2010, 2014) and Patience et al. (2012) characterized the complete 1.1–2.5 μm spectrum at medium spectral resolution with SINFONI data at the VLT. Various atomic and molecular lines were found to be present. From atmospheric and evolutionary models, a L1±1 spectral type was reported, together with a $\log(L/L_\odot) = -3.7 \pm 0.1$ dex, $T_{eff} = 1700 \pm 100$ K, and $\log(g) = 4.0 \pm 0.5$ dex.

Regarding the structure of this paper, the observations are described in Sect. 2, including the data processing for the SPHERE 2015 photometric point and the reduction methods for the K band SINFONI spectrum. In Sect. 3, we present the orbital modeling. In Sect. 4, we present the spectral modeling and provide a derivation of the best-suited physical properties. In Sect. 5, we present our estimation of the obliquity of AB Pic b and test different formation scenarios. The main conclusions of this study are outlined in Sect. 6.

## 2. Observations and data reduction

AB Pic b has been the focus of many studies since its discovery by Chauvin et al. (2005). A rich set of literature observations is available. Among them are (i) photometry points on the visible (0.53–0.92 μm) taken by the HST/WFC3 (Bonnefoy et al. 2014), (ii) medium-resolution SINFONI spectra ($R_\lambda = 1500$–2000) in the *J* and *H* bands (Bonnefoy et al. 2014), (iii) an Lp band spectrum ($R_\lambda \sim 300$) from the Magellan-AO/CLIO2 (Stone et al. 2016), and (iv) *Spitzer* IRAC photometry points at 3.6, 4.5, 5.8, and 8 μm (Martinez & Kraus 2022). Mining into unpublished archived datasets, we found medium-resolution SINFONI/VLT spectroscopic observations at *K* band from December 2013 (program 092.C − 0809 (A); PI J. Patience), which complement the spectral information at *J* and *H* bands. The system was also observed in 2015 with SPHERE at the VLT in the course of the SPHERE Guaranteed Time Observations (GTO program:

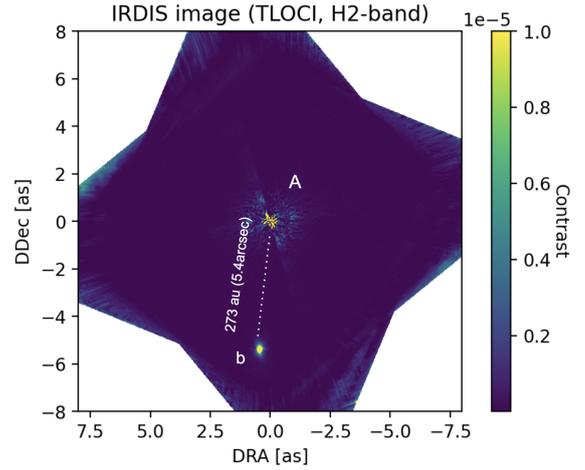

**Fig. 2.** SPHERE-IRDIS coronographic image at H band of AB Pic b relative to A reduced using SpeCal (Galicher et al. 2018).

095.C-0298 (H); PI J. L. Beuzit) to explore the orbital properties of AB Pic b in combination with previous NaCo observations from 2003 and 2004 (Chauvin et al. 2005). The new SINFONI and SPHERE observations are described below and drive this study.

### 2.1. SPHERE observations and data processing

AB Pic was observed during the SpHere INfrared survey for Exoplanets (SHINE, Guaranteed Time Observations; Chauvin et al. 2017; Desidera et al. 2021; Langlois et al. 2021; Vigan et al. 2021) on February 6, 2015, using the VLT/SPHERE high-contrast instrument (Beuzit et al. 2019). The observations were obtained with the IRDIFS mode, which combines the IRDIS (Dohlen et al. 2008) and IFS instruments (Claudi et al. 2008) simultaneously. IRDIFS combines IRDIS in dual band imaging (DBI, Vigan et al. 2010) with H2H3 filters ($\lambda_{H2} = 1.593 \pm 0.055$ μm, $\lambda_{H3} = 1.667 \pm 0.056$ μm), and IFS in the YJ (0.95–1.35 μm) setting.

All IRDIS and IFS datasets were reduced using the SPHERE Data Reduction and Handling (DRH) automated pipeline (Pavlov et al. 2008) at the SPHERE Data Center (SPHERE-DC) to correct each datacube for bad pixels, dark current, flat field, and sky background. After combining all datacubes with an adequate calculation of the parallactic angle for each frame of the deep coronographic sequence, all frames are shifted at the position of the stellar centroid calculated from the initial star center position. To calibrate the IRDIS and IFS datasets on sky, the astrometric field 47 Tuc was observed. The platescale and true north solution were extracted from the long-term analysis of the GTO astrometric calibration described by Maire et al. (2021). The rotation correction considered to align the images to the detector vertical in pupil-tracking observations is $-135.99 \pm 0.11$ deg. The anamorphism correction was obtained by stretching the image Y-direction by a factor of $1.0060 \pm 0.0002$ (inducing an error of 1.1 mas at the separation of AB Pic b, 440 pixels). AB Pic b is only seen within the IRDIS field of view (FoV), being located at more than 5.4 arcsec (see Fig. 2). Its relative position was derived using the SpeCal pipeline (Galicher et al. 2018). The final astrometric error takes into account the centering error behind the coronograph, the fitting error of the companion position given by SpeCal, and the astrometric errors coming from the anamorphism, platescale, true-north correction, and pupil-offset correction. The result is reported in Fig. 3 together with the NaCo data points from Chauvin et al. (2005). The astrometric measurements of AB Pic b used for the orbital fitting relative to AB Pic A are reported in Table 1.

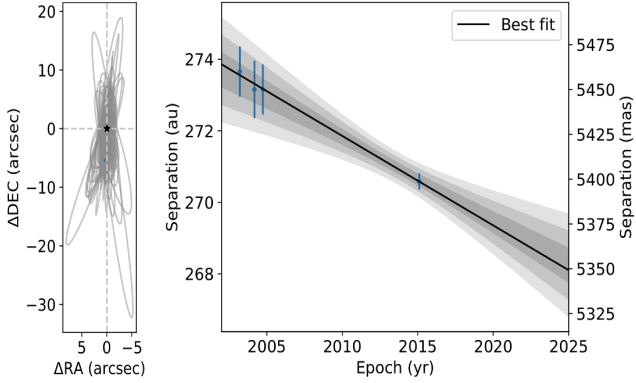

**Fig. 3.** Orbital monitoring combining NaCo and SPHERE observations. Left: representation of 100 orbits selected from the posteriors. The orbits are highly inclined (~90 deg). Right: zoom onto the orbital fitting as a function of epoch in years and projected separation. The black line represents the best orbital solution (lower $\chi^2$) and the gray areas are the 1, 2, and 3 $\sigma$ regions.

**Table 1.** Astrometric measurements of AB Pic b relative to AB Pic A.

| Telescope/instrument | Epoch | Separation (mas) | PA (deg) |
|---|---|---|---|
| VLT/NaCo | 2003.213 | 5460 ± 14 | 175.33 ± 0.18 |
| VLT/NaCo | 2004.180 | 5450 ± 16 | 175.13 ± 0.21 |
| VLT/NaCo | 2004.735 | 5450 ± 14 | 175.30 ± 0.20 |
| VLT/SPHERE-IRDIS | 2015.097 | 5398.7 ± 4.5 | 175.26 ± 0.13 |

### 2.2. SINFONI K band observations

AB Pic b was observed with SINFONI at the VLT on 13 December 2013 and 1 December 2014 (program 092.C − 0809 (A); PI J. Patience). SINFONI was composed of a custom AO module (MACAO) and an IFS (SPIFFI). SPIFFI cuts the FoV into 32 horizontal slices (slitlets), which are re-aligned to form a pseudo-slit, and dispersed by a grating on a Hawaii 2RG (2k × 2k) detector (Eisenhauer et al. 2003; Bonnet et al. 2004). The instrument was operated with pre-optics and a grating sampling of a 3″ × 3″ FoV with rectangular spaxels of 50 × 100 mas size, from 1.928 to 2.471 μm at a spectral resolution of $R_\lambda = \frac{\lambda}{\Delta\lambda} = 5090$, as mentioned in the SINFONI user manual. MACAO was used during the observations with a natural guide star (NGS) reference for wave-front sensing. A sequence of six exposures was performed of 300 s integration with five frames each day centered on the expected position of the planet. Together with the planet observations, a telluric standard star (STD) of spectral type B8V was observed each night with one exposure of 300 s.

### 2.3. SINFONI cube building and spectral extraction

We initially reduced the data with the ESO SINFONI data handling pipeline v3.0.0[3] through the EsoReflex[4] environment. The pipeline uses calibration frames to perform basic adjustments to the raw science frames and correct them for distortion. The slitlet positions on the frames at each wavelength are identified before building a datacube for each exposure. Further corrections on top of the ESO reduction steps were performed based on the Toolkit for Exoplanet deTection and chaRacterization with IfS (TExTRIS; Petrus et al. 2021; Bonnefoy et al., in prep.). A detailed description of these methods is given below.

---
[3] SINFONI Pipeline
[4] EsoReflex Software

*Wavelength calibration.* As implemented by Petrus et al. (2021), this method identifies a constant wavelength shift with respect to the telluric absorption lines. We corrected the companion and the STD from spaxel-to-spaxel wavelength shifts. The implemented method does not include an uncertainty measurement on its current version, but we observed discrepancies of up to ~15 km s⁻¹ in the recalibration in wavelength for another target.

*Spectral extraction.* We measured the motion of a point source affected by atmospheric refraction using a 2D Moffat function and recovered the center coordinates of AB Pic b and the STD in each cube as a function of wavelength. We played by adjusting a polynomial of 1 or 2 degrees and the initial guess for the expected center position of the target. For the STD frames, we binned the cubes in wavelength to reduce computation time. Different extraction radii centered at the source position were evaluated to find the optimal aperture. The spectral slope, in general, does not change with increasing aperture, and the S/N seems higher for smaller apertures; we therefore set the aperture radius to 5 pixels (~1 FWHM assuming a Gaussian distribution of the flux on the FoV). The error bars for each extracted spectrum are computed from the estimated one-standard-deviation level of the residuals around the circular aperture.

*Telluric removal.* The contamination by water bands of the Earth's atmosphere is recognizable in this preliminary spectrum. With the STD observations, we recovered an atmospheric transmission spectrum for each night, for which we first corrected the cubes from the dark spot defect of the SINFONI detector. This is a defect that is not properly interpolated by the pipeline and produces a dip in flux in the datacubes. It is known to affect the $K$ band data and we corrected it by fitting a one-dimensional polynomial from 2.14 to 2.15 μm. Next, we corrected the NIR hydrogen lines from the STD spectrum by fitting a Voigt profile. The function corrects for specific lines of the Paschen (Pa$_\beta$) and Brackett (Br₁₆, Br₁₄, Br₁₂, Br₁₁, Br₁₀, and Br$_\gamma$) series that may be observable in the NIR. We inspected these corrections, and using the spectral type of the STD star, we computed its corresponding theoretical black-body curve. Each STD spectrum was divided by the computed black body to recover the atmospheric transmission of each night. Finally, each spectrum of AB Pic b was corrected by dividing the atmospheric transmission spectrum.

*Final spectrum.* We mean-combined all observations, obtaining a final spectrum for AB Pic b with the corresponding errors, which were computed by mean-combining the errors of each spectrum and dividing this by the root of the number of datacubes considered. Before merging, we applied a Barycentric correction.

We calibrated the extracted normalized spectrum in flux units for all bands individually using the $J$, $H$, $K$, and $Lp$ band magnitude values reported in Table C.1 of Bonnefoy et al. (2014). The full spectral energy distribution (SED) for AB Pic b is observable in Fig. 4.

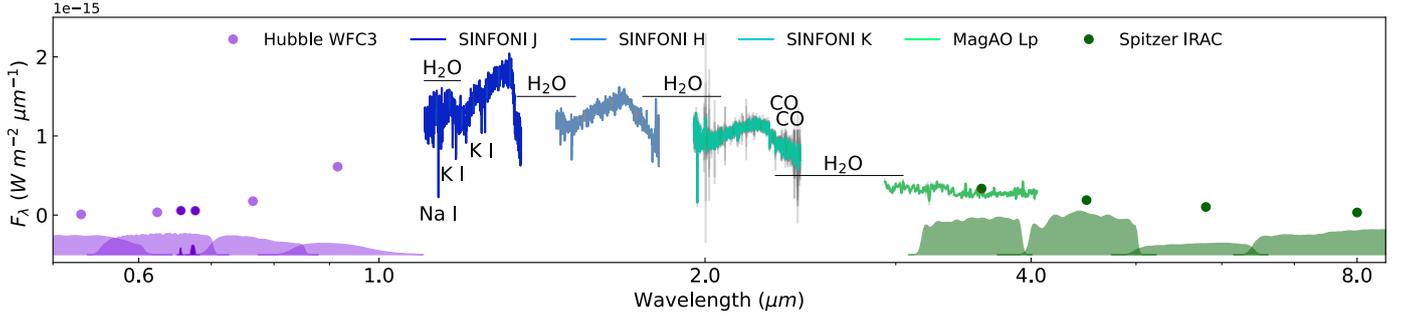

**Fig. 4.** Spectral energy distribution of AB Pic b combining photometric points taken by HST/WFC3 (*F555W*, *F625W*, *F656N*, *F673N*, *F775W*, and F850LP, listed in Table D1 from Bonnefoy et al. (2014) and *Spitzer* IRAC (1, 2, 3, and 4), with SINFONI spectra on the *J*, *H*, and *K* bands, and Magellanic AO *Lp* band. Uncertainties are represented in gray, but are too small to be noticeable for the photometric points. The most relevant and distinctive atomic and molecular absorption lines are labeled at their corresponding location. The filter profiles are represented below each photometric point.

**Table 2.** OFTI solution from NaCo and SPHERE relative astrometry.

| Orbital parameter | Prior | OFTI solution |
|---|---|---|
| $a$ (au) | 0–1000 | $190^{+200}_{-50}$ |
| $i$ (°) | 0–180 | $90 \pm 12$ |
| $\Omega$ (°) | −90–90 | $-5 \pm 13$ ($\pm 180$) |
| $T_{\rm p}$ (yr) | 0–30 000 | $2600^{+1700}_{-300}$ |

## 3. Orbital properties

The combination of NaCo observations from March 2003, March 2004, and September 2004 (Chauvin et al. 2005) with the SPHERE observation from February 2015 offers the time needed to partially resolve the orbital motion of AB Pic b. The motion is mainly resolved in separation (nothing significant is observed in position angle), which indicates a highly inclined orbit for the planet. We used the Orbits for the Impatient (OFTI) tool – a Bayesian rejection sampling method for quickly fitting the orbits of long-period exoplanets (Blunt et al. 2017) – to explore the solutions. We adopted the default priors detailed in the *'orbitize!'* Python package (Blunt et al. 2020), which is logarithmic for the semi-major axis and flat for the eccentricity. We set an upper bound of 1000 au for the semi-major axis because solutions greater than 1000 au have a probability that is 100 times lower than the most likely value. The best solution is observable in Fig. 3 and detailed in Table 2. We note that AB Pic b was directly detected by *Gaia*, but we decided to not use the *Gaia* eDR3 data point given the poor astrometric fitting solution due to the faintness of AB Pic b in the optical and its close separation. The corner plot of the orbital parameters is shown in Fig. A.1. The solution confirms that the exoplanet AB Pic b is likely to be orbiting at a semi-major axis of a = $190^{+200}_{-50}$ au on a highly inclined orbit ($i = 90 \pm 12$ deg), with a potentially moderate to high-eccentricity, albeit poorly constrained. As expected by the north–south orientation of the orbit, the longitude of the ascending node ($\Omega$) is close to ~0 deg ($\pm 180$ deg).

## 4. Physical properties

We followed two approaches to derive the physical properties of AB Pic b. We first used the BEX-Hottest-cond03 evolutionary model (Marleau et al. 2019) for predicting the $R$, $T_{\rm eff}$, and log($g$). We then carried out a forward modeling analysis with ForMoSA (short for Forward modeling tool for Spectral Analysis), which is presented and detailed in Petrus et al. (2020, 2021). In ForMoSA, we implemented two different atmospheric models, the 2013 version of the BT-SETTL model grid exploring different $^C/_O$ ratios (Allard et al. 2013) and Exo-REM Charnay et al. (2021) in order to analyze their performance and limitations at deriving robust physical properties.

### 4.1. Evolutionary model predictions

In Fig. 5, we show the evolutionary tracks of the BEX-Hottest-cond03 models (Marleau et al. 2019) as a function of $R$, log($g$), age, and $T_{\rm eff}$. Considering an age of $13.3^{+1.1}_{-0.6}$ Myr (Booth et al. 2021), the distance of $50.14 \pm 0.04$ pc to the Sun, and the NaCo $J$, $H$, and $K$ magnitudes, we predicted the physical properties of AB Pic b for different bands (see Table 3). We sample from the evolutionary models considering the asymmetric uncertainties in age. However, for simplicity, the $R$, log($g$), and $T_{\rm eff}$ were computed assuming a symmetric distribution of the selected subsample. The evolutionary models predict for this target, based on the NaCo photometric points, $R = 1.53 \pm 0.03\,R_{\rm Jup}$, log($g$) = $4.17 \pm 0.02$ dex, and $T_{\rm eff} = 1855 \pm 30$ K, which are consistent with Bonnefoy et al. (2014).

### 4.2. Atmospheric forward-modeling predictions

To explore the spectral diversity of the photometric and spectroscopic observations we used ForMoSA, a tool based on a forward-modeling approach that compares observations with grids of pre-computed synthetic atmospheric models using Bayesian inference methods (see Petrus et al. 2020, 2021). ForMoSA relies on a nested sampling algorithm (Skilling 2006) to determine the posterior distribution function (PDF) of a set of free parameters. This method performs a global exploration to look for local maxima of likelihoods, iteratively isolating a progressively restrained area of the same likelihood while converging toward the maxima. It avoids missing local maxima of likelihood because it can encapsulate different regions. For-MoSA generates synthetic spectra at each step of the nested sampling using linear interpolation of the spectra from the original model grid. The model selection relies on the Bayesian evidence (Trotta 2008). The errors given by ForMoSA are statistical and were determined for each parameter as the range that encompasses $1\sigma$ assuming Gaussian distributions, and they do not include possible systematic errors in the models (Petrus et al. 2021).

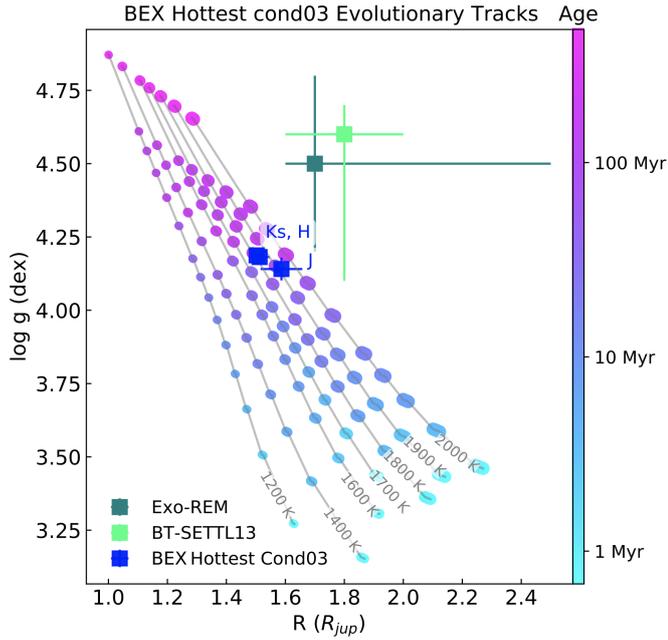

**Fig. 5.** Radius vs. gravity as a function of age and effective temperature from the BEX Hottest cond03 evolutionary models. The color scale traces the ages on a logarithmic scale, and iso-temperature curves are labeled. The estimated position in the diagram for AB Pic b is reported for (i) the evolutionary model prediction (blue) from the BEX Hottest cond03 models using the absolute magnitudes from NaCo in the *J*, *H*, and *K* band (Chauvin et al. 2005) and (ii) the ForMoSA predictions (green) from our adopted values for both families of atmospheric models implemented (see Sect. 4.3).

**Table 3.** AB Pic b predictions from BEX-Hottest cond03 models.

| Band | Magnitude | $R$ ($R_{\rm Jup}$) | $\log(g)$ (dex) | $T_{\rm eff}$ (K) |
|---|---|---|---|---|
| *J* | $12.27 \pm 0.36$ | $1.57 \pm 0.07$ | $4.14 \pm 0.04$ | $1914 \pm 75$ |
| *H* | $11.38 \pm 0.18$ | $1.51 \pm 0.03$ | $4.18 \pm 0.02$ | $1833 \pm 45$ |
| *K* | $10.59 \pm 0.07$ | $1.50 \pm 0.02$ | $4.19 \pm 0.01$ | $1820 \pm 21$ |

**Notes.** These predictions are for the calibrated NACO absolute magnitudes on the *J*, *H*, and *K* bands.

ForMoSA, in general, allows the user to fit the atmospheric grid for $T_{\rm eff}$ and $\log(g)$, and [M/H] and $^C/_O$ ratio may be included, among others, depending on the model. In addition, the RV, the radius of the planet (R), the extinction coefficient ($A_v$), the rotational velocity ($v\sin(i)$), and the limb-darkening ($\epsilon$) can be included as free parameters. When including them, the models are modified accordingly before being compared to the observations. The allowed ranges to explore the solutions that we considered are listed in the first row of Table B.1 for BT-SETTL13 and Table B.2 for Exo-REM. We explored the limb darkening ($\epsilon$) from 0.01 to 0.99 every time we included the $v\sin(i)$ in the models, because the former is necessary to compute the latter. However, this is a parameter used for resolved planets. In our case, its posterior has no physical meaning because we do not know its distributions for brown dwarfs and exoplanets. Therefore, we exclude the limb darkening posteriors from Tables B.1 and B.2, but four examples are shown in the corner plots of Figs. 7 and 8.

### 4.2.1. Atmospheric models

**BT-SETTL13.** This model is part of a family of atmospheric models by Allard et al. (2013) that include 1D clouds where the abundance and size distribution of 55 types of grain are determined by comparing the timescale of condensation, coalescence, gravitational settling, and mixing of solids, with these latter assumed to be spherical. The details of each solid and chemical element are described in Rajpurohit et al. (2018). The opacities are calculated for each line and the radiative transfer simulations are carried on by the PHOENIX code (Hauschildt et al. 1997; Allard et al. 2001). The convection is handled following mixing-length theory and works at hydrostatic and chemical equilibrium. The model also accounts for non-equilibrium chemistry between CO, $CH_4$, $CO_2$, $N_2$, and $NH_3$. The grid we used provides spectra from 0.3 to 15 µm and considers $T_{\rm eff}$ from 1400 to 2200 K, $\log(g)$ from 3.5 to 5.0 dex, $^C/_O$ ratio from 0.2754 to 1.096, and a constant solar metallicity ([M/H] = 0.0). In ForMoSA, we interpolate the grid to the resolution of the observations. The free parameters are linearly interpolated to perform the exploration. The posteriors, obtained using 500 living points, are reported in Table B.1.

**Exo-REM.** This model is an atmospheric radiative–convective equilibrium model presented in Baudino et al. (2015), Charnay et al. (2018, 2021), and Blain et al. (2021) including a cloud description well suited to reproducing spectra where dust dominates, especially at the L-T transition. Like the BT-SETTL13 models, the atmosphere is cut into pressure levels where the flux is calculated iteratively using a constrained linear inversion method for radiative–convective equilibrium. The initial abundances of each chemical element are first established using the values tabulated in Lodders (2010). The model includes the collision-induced absorption of $H_2$–$H_2$ and $H_2$–He, ro-vibrational bands from nine molecules ($H_2O$, $CH_4$, CO, $CO_2$, $NH_3$, $PH_3$, TiO, VO, and FeH), and resonant lines from Na and K. As in BT-SETTL13, Exo-REM accounts for nonequilibrium chemistry between CO, $CH_4$, $CO_2$, and $NH_3$ due to vertical mixing. The abundances of the other species are computed at thermochemical equilibrium. The vertical mixing is parametrized by an eddy mixing coefficient from cloud-free simulations. The model includes iron and silicate clouds. This grid provides spectra from 0.6667 to 251.6 µm and includes four free parameters: $T_{\rm eff}$ from 400 to 2000 K, $\log(g)$ from 3.0 to 5.0 dex, $^C/_O$ ratio from 0.1 to 0.8, and [M/H] from −0.5 to 1.0. When implementing the Exo-REM models, we excluded the three photometric points below 0.67 µm from the fit because of the wavelength range of this grid. Those missing photometric points do not impact the posteriors, given their low constraining power. This can be verified when comparing models B0 and B2; excluding all the photometric points leads to a $\Delta T_{\rm eff} \sim 1$ K.

### 4.2.2. Fitting strategy

For the forward-modeling approach, we explored fitting different SED regions and parameters. See Tables B.1 and B.2 for the labels of each model. Our strategy is the following:

i. We first considered the full available SED (*J*, *H*, *K*, and *Lp* bands with and without photometric points). Here, we expect a reliable $T_{\rm eff}$ estimation given the broad spectrum range modeled, where the black-body behavior is identifiable.

ii. Second, we analyzed the *K* band. A good fit of the CO bandheads is expected to derive a reliable $^C/_O$ ratio. We tried fitting different extra parameters and subtracting the continuum by computing a low-resolution spectrum ($R \sim 100$) and subtracting it from the observations and models.

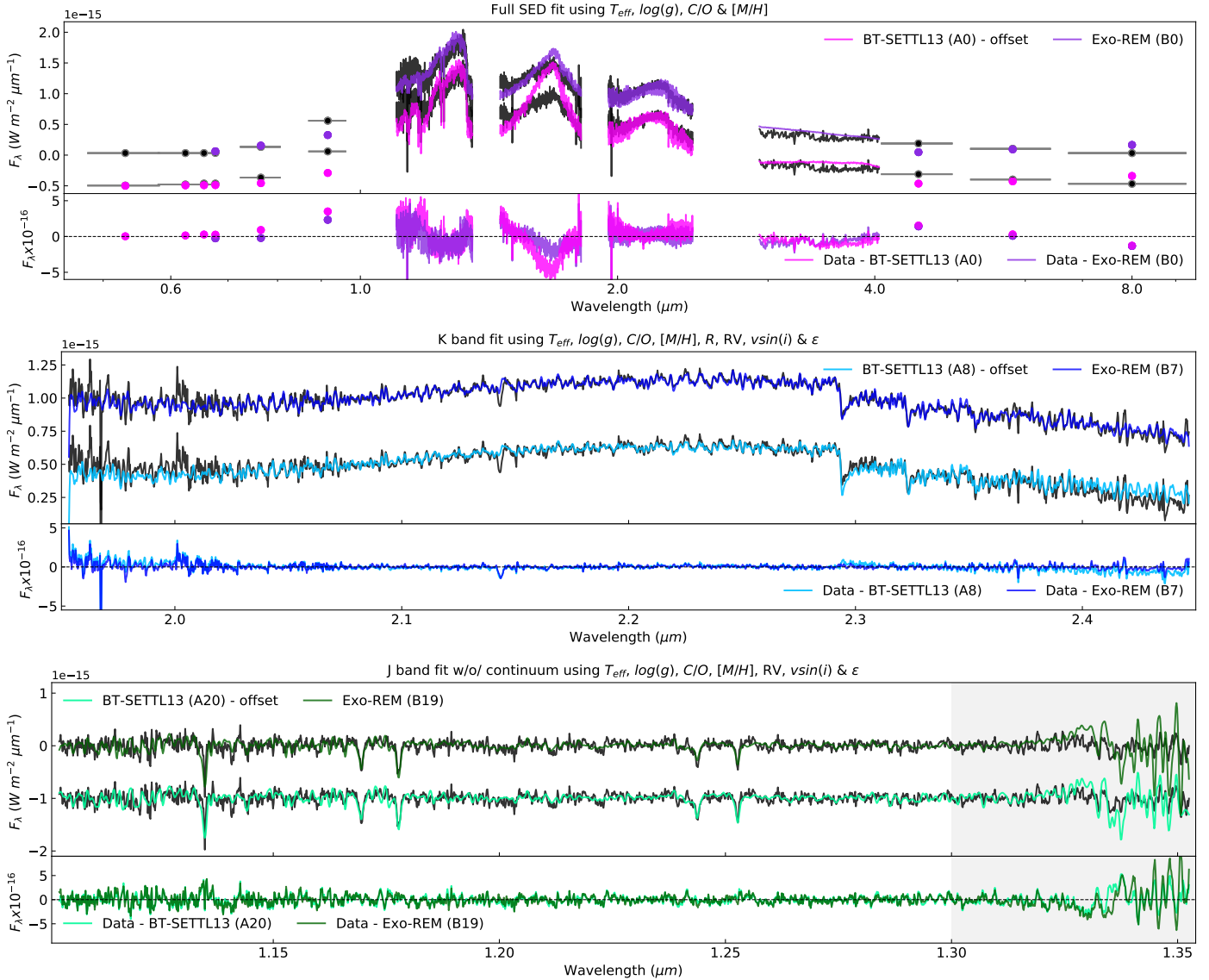

**Fig. 6.** Best atmospheric models. *Top*: full SED fit for both families of atmospheric models and the (data–model) residuals. An offset was applied to the BT-SETTL3 model. *Middle*: same as the top but for the *K* band only. *Bottom*: same as the top but for the *J* band modeled without the continuum. Here, the gray area (> 1.3 μm) shows the spectral region excluded from the fit. Each model is identifiable by the label with the numeration of Tables B.1 and B.2. The colors are consistent with the corner plots (Figs. 7 and 8) and the tables.

iii. Third, we considered the *J* band. This band has K I absorption lines that have great potential for deriving reliable $\log(g)$ (Gorlova et al. 2003; Allers & Liu 2013; Bonnefoy et al. 2014). Due to the $H_2O$ bands from 1.3 to 1.55 μm, we restricted the wavelength range for the nested sampling up to 1.3 μm. For the *K* band, we tried different fittings and subtracting the continuum.

iv. Finally, *H* and *Lp* bands were modeled with the continuum, but mainly as an exploratory exercise given the low resolution for the Lp band and the poor *H* band fit from both BT-SETTL3 and Exo-REM full SED models.

For BT-SETTL3, three examples from the several models tested are observable in Fig. 6 under the labels A0 (full SED), A8 (*K* band with continuum), and A20 (*J* band without continuum). The posterior distributions are shown in Fig. 7 together with the adopted values (dashed black lines) that were selected for the reasons outlined above; these are also listed in Table 4.

We followed the same approach for Exo-REM. Three examples from these atmospheric models are observable in Fig. 6 under the labels B0 (full SED), B7 (*K* band with continuum), and B19 (*J* band without continuum), the posteriors are shown in Fig. 8, and the adopted values in Table 4.

### 4.3. Final adopted values

Regarding the final physical properties of AB Pic b, in general, both atmospheric models performed well at first order, but a careful look into each parameter is essential given the multi-modeling approach and current model limitations. The results of the models are classified by the spectral range and specific free parameters included (see Tables B.1 and B.2). Previous works by Petrus et al. (2020) showed that key limitations when extracting the physical properties of young exoplanets, and more broadly, when interpreting the results in terms of the processes at play in young exoplanetary atmospheres, come from the atmospheric

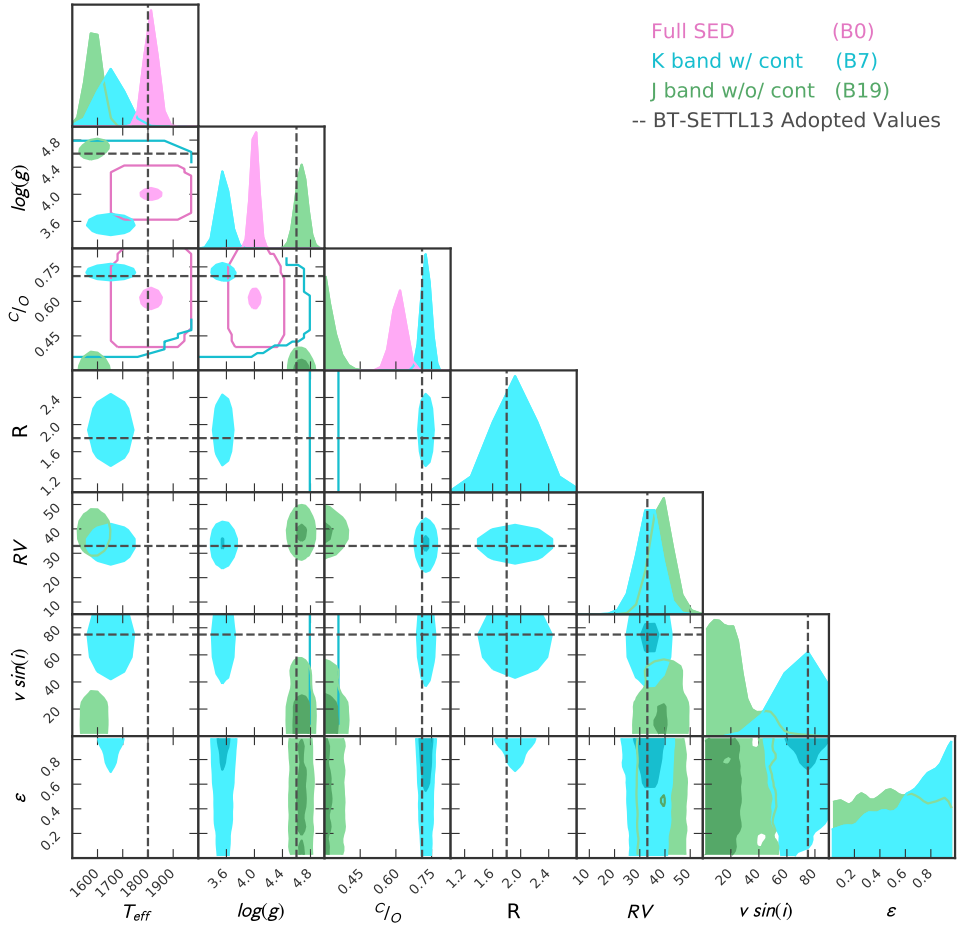

**Fig. 7.** Corner-plot comparing the posteriors of the models presented in Fig. 6 for BT-SETTL13 with consistent colors. The black dashed lines represent the adopted values from this work listed in Table 4.

**Table 4.** Final adopted atmospheric physical properties from ForMoSA.

| Parameter | BT-SETTL13 | Exo-REM |
|---|---|---|
| $T_{eff}$ (K) | $1800 \pm 20$ | $1700 \pm 50$ |
| $\log(g)$ (dex) | $4.6^{+0.1}_{-0.5}$ | $4.5 \pm 0.3$ |
| [M/H] | 0.0 | $0.36 \pm 0.20$ |
| $^C/_O$ | $0.71 \pm 0.15$ | $0.58 \pm 0.08$ |
| $R$ ($R_{Jup}$) | $1.8 \pm 0.2$ | $1.7^{+0.8}_{-0.1}$ |
| RV (km s$^{-1}$) | $33 \pm 10$ | $32 \pm 6$ |
| $v\sin(i)$ (km s$^{-1}$) | $75^{+200}_{-16}$ | $73^{+11}_{-27}$ |

models themselves. This is particularly true when high signal-to-noise ratio medium-resolution spectra at $J$, $H$, and $K$ bands are available, as in the case of AB Pic b. For a given family of models, the forward-modeling statistical fitting error of the extracted bulk properties is relatively small, although the solutions generally fail to perfectly reproduce the pseudo-continuum over a broad range of wavelengths. The use of different spectral ranges, each sensitive to different physical parameters (effective temperature, surface gravity, metallicity, C/O), combined with different families of atmospheric models, is an interesting and pragmatic strategy to explore the range of physical solutions that emerge from the models, considering their different prescriptions (temperature structure, thermo-compositional convection, and instabilities, clouds, and opacities etc.). We also report the

Bayesian evidence in Tables B.1 and B.2 which can be used to compare the results of both models for the same set of fitting boundaries. We summarize the adopted values using both BT-SETTL13 and Exo-REM below.

From the full SED fitting, we derive $T_{eff} = 1800 \pm 20$ K for BT-SETTL13, and $T_{eff} = 1700 \pm 50$ K for Exo-REM. These temperatures were selected from the full SED fit models A0 and B0. Both values are consistent with the previous estimation by Bonnefoy et al. (2014), and with the evolutionary model predictions. These two values do not strictly agree at a $1\sigma$ confidence level. This may stem from a systematic difference in the representation of the overall spectral continuum by the two sets of models, which affects the temperature determination and is not taken into account in the error bars.

For the surface gravity, as mentioned, the K I lines are particularly interesting. Both atmospheric models have difficulty in performing a good fit of the $J$ band continuum, which is probably due to the poorly reproduced $H_2O$ signatures. If we remove the continuum and restrict the wavelength range, we derive $\log(g) = 4.6^{+0.1}_{-0.5}$ dex with BT-SETTL13, which is slightly higher than the predictions of the evolutionary models and the reported value from Bonnefoy et al. (2014), although within the error bars. This overestimation of $\log(g)$ with BT-SETTL13 is a global problem encountered for a wider range of young L- and T-type brown dwarfs (Petrus et al., in prep.). The Exo-REM models exhibit the same continuum $J$ band issue. However, as this family of models allows exploration of [M/H], the posteriors are slightly harder to interpret. From the $J$ band without continuum, we derive a [M/H] ∼ −0.36 for high log($g$) ∼ 4.5 dex. This [M/H] value is highly doubtful physically. When analyzing these two parameters in the $K$ band models, we observe a correlation between log($g$) and [M/H]. The best fit for the $K$ band (B7) derives a [M/H] = 0.36 ± 0.20 and log($g$)∼4 dex; however, higher [M/H] values are obtained when removing the continuum. By comparing with the evolutionary models, we observe that the most consistent results are the ones with lower [M/H] and higher log($g$), and therefore our adopted values are log($g$) = 4.6$^{+0.1}_{-0.5}$ dex for BT-SETTL13, and for Exo-REM the mean value between the B7 and B19 models, leading to [M/H] = 0.36 ± 0.20 and log($g$) = 4.5 ± 0.3 dex. The [M/H] is not a parameter we can rely on because it is highly degenerate with other parameters, such as the overall shape, the flux calibration of the SED, and log($g$).

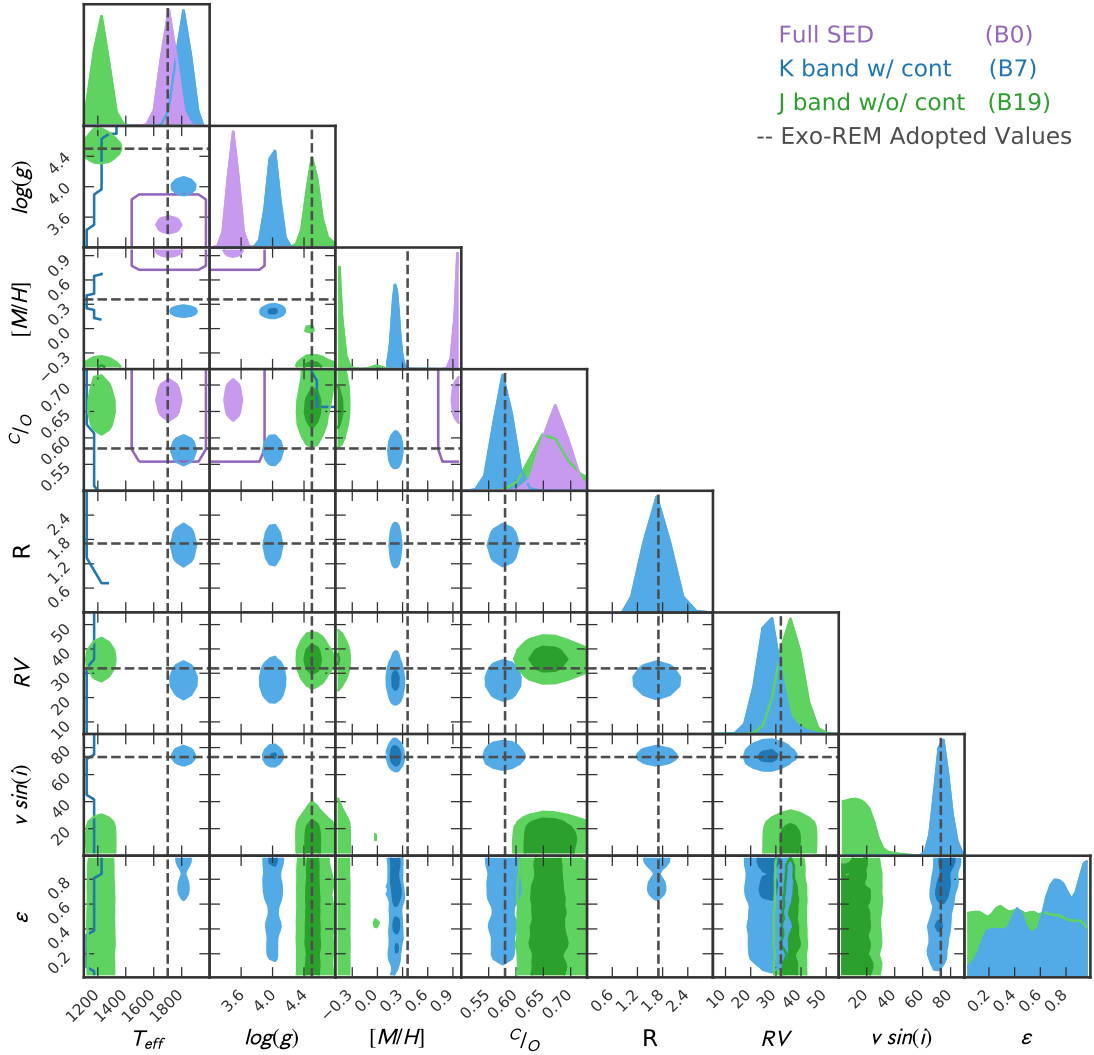

**Fig. 8.** Same as Fig. 7 but for Exo-REM models.

For the determination of the $^C/_O$ ratio, when we zoom onto the CO bandhead region in the $K$ band (see Fig. 9), we observe that the atmospheric model B7, with $^C/_O$ = 0.58 ± 0.08, matches the data exquisitely. For models with higher $^C/_O$ ratios, we start observing more CO lines than the ones we resolve in the data and for models with lower values we do not recover the overall shape of the CO bandheads; models B3 and B16 in Fig. 9 illustrate this behaviour. In general, a model with a stellar $^C/_O$ ratio is in better agreement with the data. In this line, the Exo-REM $K$ band model derives a precise prediction of the continuum shape in addition to the good fit over the CO bandheads (see model B7 in the middle panel of Fig. 6 compared to model A8). In conclusion, we adopt a $^C/_O$ ratio of 0.58 ± 0.08.

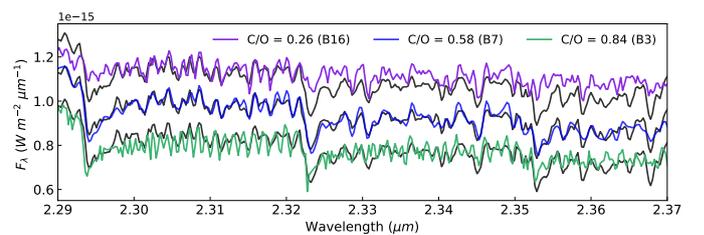

**Fig. 9.** Zoom onto the CO bandheads modeled with Exo-REM labeled with their $^C/_O$ value. Black curves are the observations.

For both atmospheric models, the $K$ band with continuum derived the most confident radius posterior, namely 1.8 ± 0.2 $R_{Jup}$ for BT-SETTL13 and 1.7$^{+0.8}_{-0.1}$ $R_{Jup}$ for Exo-REM. We rely on the estimation of this setup because the modeled spectrum better fits all the spectral features (see Fig. 6). This value is slightly higher than the one reported by Bonnefoy et al. (2014) of 1.4 ± 0.3 $R_{Jup}$, and the predictions of the evolutionary models of 1.53 ± 0.03 $R_{Jup}$, but is almost within the error bars. The reason for our bigger radius could be a degeneracy with another parameter, such as $T_{eff}$, or an incorrect prediction of the emergent flux by the models. Even though the radius should be linked to the $T_{eff}$, we avoided estimating it from the full SED models because of the possibly biased flux calibration over the different wavelength ranges.

For the RV, we report a value of $33 \pm 10 \, \text{km s}^{-1}$ with BT-SETTL13 and of $32 \pm 6 \, \text{km s}^{-1}$ with Exo-REM, obtained by averaging the $K$ band without continuum posteriors and considering the confidence intervals. This value is derived from the $K$ band models because this band was re-calibrated in wavelength, which is necessary for a reliable outcome, and has a higher spectral resolution. Additional uncertainties of up to $15 \, \text{km s}^{-1}$ are expected, coming from the recalibration in wavelength, which is not taken into account by ForMoSA. With high-resolution data, we could constrain the RV with a precision of down to $3 \, \text{km s}^{-1}$ (see Wang et al. 2021a), allowing us to confirm that AB Pic b is physically bound to AB Pic A. The $v\sin(i)$ has an erratic behavior for the different wavelength ranges, and is probably biased by the other free parameters. However, given the precise line fitting of the B7 model, we trust the value of $73^{+11}_{-27} \, \text{km s}^{-1}$. Finally, for $A_v$ we see very different behaviors, but this has previously been reported to take a relatively low value ($A_v \sim 0.2$) by Bonnefoy et al. (2014). This parameter is expected to lead to reddening of the observations due to interstellar dust, but given the proximity of AB Pic ($d = 50.14 \pm 0.04$ pc), this value is expected to remain relatively low as found in the forward-modeling analysis. The final adopted values for AB Pic b are reported in Table 4.

## 5. Discussion

### 5.1. Obliquity of the planet

Using the derived $v\sin(i) = 73^{+11}_{-27} \, \text{km s}^{-1}$ from the B7 model, the orbital inclination of the planet ($i_o$) of $\sim 90$ deg from the orbital fitting, and the reported rotational period of $\sim 2.1$ h from Zhou et al. (2019), we can set some constraints on the inclination of its spin axis, and its true obliquity ($\Psi_{op}$). We followed the approach described by Bryan et al. (2021; Sect. 3.5 for the HD 106906 b case) derived from Masuda & Winn (2020), which takes into account the fact that $v$ and $v\sin(i)$ are not statistically independent given that $v\sin(i)$ is always less than $v$. The determination of the probability distribution for vsin(i) from our observations, and for $v$ by incorporating uncertainties on $P_{rot}$ and $R$ in a Monte Carlo fashion, lead to the determination of the posterior distribution function (PDF) for the spin axis inclination of the planet ($i$), hereafter referred to as $i_p$. This is done under the assumption that the data sets from which the likelihood functions are calculated are independent, and that $v$ and $i_p$ are a priori independent. The resulting $i_p$, together with the planet's orbital inclination ($i_o$), are reported in the top panel of Fig. 10. We already observe a significant difference between the planet's orbital plane orientation and its spin orientation in the projected celestial plane.

Considering the 3D configuration of the system, we can derive the true companion obliquity given by:

$$\Psi_{op} = \cos^{-1}(\cos(i_p)\cos(i_o) + \sin(i_p)\sin(i_o)\cos(\Omega_o - \Omega_p)). \quad (1)$$

In our case, the nodal angle $\Omega_p$ is unknown. We therefore assume values randomly drawn from a uniform distribution between 0 and $2\pi$. The resulting PDFs for the companion's line-of-sight-projected obliquity ($|i_p - i_o|$) and its true obliquity ($\Psi_{op}$) are shown in the bottom panel of Fig. 10. Interestingly, relatively high values are found for the obliquity of AB Pic b, ranging between 45 and 135 deg, indicating a rather important misalignment between the planet's spin and orbit orientations.

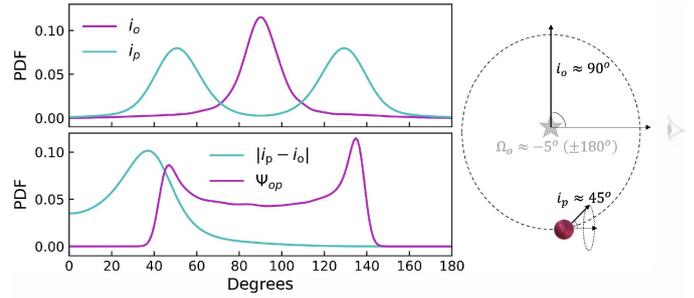

**Fig. 10.** *Top*: probability density function (PDF) for the inclination of the orbit of AB Pic b ($i_o$) together with its spin axis inclination ($i_p$). *Bottom*: PDF for the companion's line-of-sight-projected obliquity ($|i_p - i_o|$) and the true companion obliquity ($\Psi_{op}$). *Right*: architecture of the AB Pic system.

It is worth mentioning that the rotational period of 2.1 h reported by Zhou et al. (2019) is significantly shorter than the ones derived for other directly imaged planetary mass companions, as pointed out by Xuan et al. (2020). Its rotational velocity is also significantly higher and close to $\sim 70\%$ of the break-up velocity ($v_{break-up} = 102 \, \text{km s}^{-1}$), compared to values of around 5–30% for other planetary-mass companions. Confirming the planet's true obliquity and rotational period, together with a better constraint on the orbital eccentricity, would be important in the near future to confirm the singularity of AB Pic b, and shed light on its dynamical history given its current location, in connection with the presence of possible inner companion(s).

### 5.2. A possible second, inner planet

To explore the possible presence of additional inner planets in the system, we explored the completeness of the new SPHERE high-contrast imaging observation using the MESS2 code (Lannier et al. 2017), an evolution of the Multi-purpose Exoplanet Simulation System (MESS), a Monte Carlo tool for the statistical analysis of direct imaging survey results (Bonavita et al. 2012). MESS2 combines the information on the star with the instrument detection limits to estimate the probability of detection of a given synthetic planet population, ultimately generating detection probability maps. The code generates a grid of masses and physical separations of synthetic companions, and then estimates the probability of detection given the provided detection limits. By default, all the orbital parameters are uniformly distributed except for the eccentricity, which is generated using a Gaussian eccentricity distribution with $\mu = 0$ and $\sigma = 0.3$ (constraint: $e \geq 0$), following the approach by Hogg et al. (2010). This allows us to properly take into account the effects of projection when estimating the detection probability using the contrast limits. The result for the case without any prior regarding the system configuration is reported as blue probability areas labeled "DI" (direct imaging) in Fig. 11.

To further explore the presence of additional planets in the system, we searched for the existence of a proper motion anomaly (PMa) based on the compilation of Kervella et al. (2022). The long time baseline of 24.75 yr between the HIPPARCOS and *Gaia* Early Data Release 3 (eDR3) position measurements is exploited to pinpoint the presence of a secondary orbiting body in the system. For AB Pic, the tangential velocity anomaly is $\Delta v_{tan} = 65.45 \pm 10.46 \, \text{m s}^{-1}$ with a $S/N = 6.3$. As discussed by Kervella et al. (2019), the mass of the companion exhibiting a PMa signal can be constrained using the measured tangential velocity anomaly. However, this relation is degenerate with the orbital radius. We plotted this prediction in pink over the "DI" probability areas in Fig. 11.

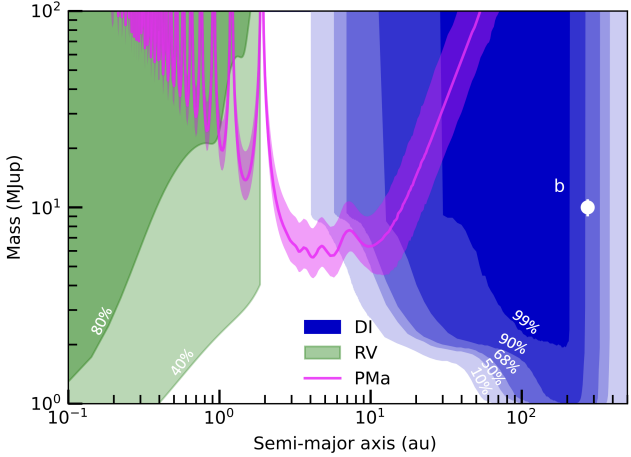

**Fig. 11.** Mass detection limits for an inner planet without assuming coplanarity with AB Pic b. The pink curve shows the estimated mass and orbital distance from the HIPPARCOS–*Gaia* proper motion anomaly (PMa), the blue area represents the SPHERE detection limits (DI), and the green area represents the RV detection limits from Grandjean et al. (2020) to give a rough idea of the inner regions that we can exclude. The semi-major axis and mass of AB Pic b are show by the white point.

We notice that AB Pic b cannot explain the PMa signal of AB Pic given its current mass and projected distance. At least one additional body must be invoked. The SPHERE detection limits exclude the presence of an additional companion down to ~10 au with a mass of larger than 2 $M_{Jup}$ excluding the most massive and long-orbit solutions compatible with the PMa. Grandjean et al. (2020) analyzed the RV signals of a sample of young solar-type stars including AB Pic, but did not find any sign of binarity or close-in brown dwarf or massive planetary mass companions in their measurements. Using the survey completeness reported by these authors (see their Fig. 12), converted from period to semi-major-axis, we show in Fig. 11 that the most massive stellar and brown dwarf solutions consistent with the PMa signal can be almost all excluded, suggesting the presence of an additional inner massive planet (AB Pic c) with a mass of ~6 $M_{Jup}$ orbiting between 2.5 and 10 au (40–200 mas). A preliminary analysis considering the astrometric noise of the *Gaia* measurements will be published as part of a detailed analysis combining the detection limits of the three observing techniques mentioned above.

### 5.3. Formation pathways of AB Pic b

Our analysis of the NaCo and SPHERE astrometry confirms that AB Pic b is orbiting at a relatively wide separation from the young, solar-type star AB Pic on a very inclined orbit, which is poorly constrained in terms of eccentricity. Our RV measurement does not currently offer the possibility to accurately constrain the relative RV to the primary host. This configuration leaves various possibilities in terms of interpretation for the formation and evolution of the planet. Its current location at ~270 au is too far away to be explained directly by a standard CA scenario (Chabrier et al. 2014). Outward migration mechanisms (planet–planet scattering, stellar fly-by) would be essential to place the planet, if formed closer-in, at such a large separation beyond the CO snow line (Marleau et al. 2019). Our current constraints on the eccentricity distribution are too loose to pinpoint some past dynamical interactions with an inner companion. From the detection limits (see Fig. 11), we show that the presence of an inner massive companion to AB Pic b with a mass $\gtrsim 2\,M_{Jup}$ at a physical projected separation $\gtrsim 10$ au is very unlikely. Although the unseen ~6 $M_{Jup}$ exoplanet orbiting at less than 10 au (Kervella et al. 2022) must be confirmed, this would rather support a loose multiple-planetary architecture as seen for other systems, such as TYC-8998-760-1 (Bohn et al. 2020), and potentially also proposed for HIP 65426 (Marleau et al. 2019).

A formation through the GI scenario (Paardekooper & Johansen 2018) cannot be excluded either and would better fit with the planet's current location, but, as for CA, this formation scenario would also require that the planet-forming disk surrounding the host star be very massive and extended at early stages. Currently, this aspect is at odds with the fact that there is no large infrared excess known for AB Pic. Zuckerman et al. (2011) reported no detection of excess at 8 μm or 70 μm with *Spitzer*, but evidence for excess emission in both the 12 μm IRAS (~25 %) and 24 μm MIPS (~75 %) channels. This emission could indicate the potential presence of a warm belt, but suggest that there is a rather very limited amount of circumstellar material left. Given the mass, age, and position of the possible companions, the circumstellar disk is expected to have already dissipate, which is consistent with observations.

Finally, a brown dwarf formation process through gravoturbulent fragmentation in the molecular cloud also seems feasible, particularly if the inner HIPPARCOS-*Gaia* planet is not confirmed, although there is still very little known about the formation of these extreme mass–ratio binaries (Nayakshin 2017).

Looking more at the revisited atmospheric properties of AB Pic b, our analysis corroborates the previous determinations of $T_{eff}$, $\log(g)$, and $R$. For the first time, we have estimated the $C/O$ ratio and the [M/H] for this planet. At this stage, we cannot derive firm conclusions for the metallicity ([M/H] = 0.36 ± 0.20) given the large uncertainties; but a value higher than the host star ([M/H]$_\star$ = 0.04 ± 0.02, Swastik et al. 2021) could potentially support a formation by CA. Even though there is no available measurement for the $C/O$ ratio of AB Pic A, the $C/O$ ratio for AB Pic b of $C/O = 0.58$ is compatible with the ISM evolution (Chiappini et al. 2003) and solar values ($C/O_\odot = 0.55$, Brewer & Fischer 2016), considering that AB Pic system is younger than our solar system. However, this $C/O$ ratio is higher than the measurements for β Pictoris b, and HIP 65426 b. β Pictoris b has a $C/O = 0.43 ± 0.05$ and given its location and likelihood of having accreted solids between the $H_2O$ and $CO_2$ snow lines, CA is the favored formation scenario, as described by Nowak et al. (2020). HIP 65426 b has a reported upper limit of $C/O \leq 0.55$ and a CA formation scenario was proposed as well by Petrus et al. (2021) given its similarity to β Pictoris b, albeit scattered later on at a larger separation through dynamical interactions. As discussed before, AB Pic b could be the product of a similar formation and dynamical process, although the existence of the inner HIPPARCOS-*Gaia* planet remains to be confirmed.

Consequently, given our current knowledge of the atmospheric and orbital properties of AB Pic b, and of the system's architecture, AB Pic b could have been formed:

i. beyond the CO snow line, as there is no clear enrichment of the abundance of either of these two atoms compared to the ISM.

In this case, it could have formed by GI in a planet-forming disk or even by gravoturbulent fragmentation at the early phase of stellar-host formation. Its current location would be likely close to its location of formation;

ii. inside the $H_2O$ snow line probably by CA, before being scattered outward to wider orbits in a similar way to that described in the mechanism proposed for HIP 65426 b. The proper motion anomaly seen by HIPPARCOS–*Gaia* indicates the potential existence of an inner unseen exoplanet below 10 au that could be responsible for this planet–planet scattering event;

iii. somewhere in between, by CA or GI, having lowered its $^C/_O$ ratio by the accretion of solids at final formation stages, following the ideas of Madhusudhan (2012). Migration mechanisms are required in this scenario as well. For both CA and GI, the question of the small amount of IR excess detected, and therefore the lack of residual material left over from the original planet-forming disk, remains to be explained.

### 5.4. Prospects

A first follow-up work on the system would be to confirm the presence of AB Pic c, the inner HIPPARCOS–*Gaia* planet. As shown in Fig. 11, directly imaging it is very challenging for current instruments, but high-precision RV measurements for the host star could help to unveil it.

In addition, the system's edge-on configuration makes it very favorable for future high-precision RV measurements on AB Pic b, as previously done with CRIRES for $\beta$ Pic b and GQ Lupi b, and more recently with KPIC for HR 8799 bcde (Snellen et al. 2014; Schwarz et al. 2016; Wang et al. 2021b), and with Gemini/IGRINS for HD 106906 b (Bryan et al. 2021). In conjunction with NaCo and SPHERE relative astrometry, this would help to obtain the three-dimensional position and velocity of AB Pic b relative to A, which has an RV = $22.65 \pm 0.04$ km s$^{-1}$ (Soubiran et al. 2018). A precise measurement of the RV could provide confirmation that the planet is physically bound to AB Pic A by a direct comparison of the projected velocity of the companion with the system escape velocity ($3.3 \pm 0.2$ km s$^{-1}$; using a separation of 190 au and a mass of 1.3 $M_\odot$). Moreover, we could directly measure the relative RV with the primary to constrain the planet's orbital eccentricity. This would enable us to confirm the rotational period of ~2.1 h, and the planet's spin axis orientation of ~40 deg. The spin result also provides information about the initial angular momentum gained at formation (Bryan et al. 2021).

Access to higher S/N observations at medium and/or high spectral resolution might also help to measure the abundances of the carbon isotopologs. AB Pic b is very similar in terms of its physical and chemical characteristics to the wide-orbit planets TYC 8998-760-1 b and c (Bohn et al. 2020). Therefore, measuring the $^{12}CO/^{13}CO$ ratio as done by Zhang et al. (2021) is a natural next step to explore the accretion of carbon ices by the planet and better constrain its history of formation and evolution.

Finally, AB Pic b will be a prime target for JWST, as demonstrated in Fig. 12 where we show the predicted spectrum based on close values to our derived atmospheric properties. JWST will observe from 0.6 up to 28 µm with the NIRSpec and MIRI spectroscopic modes. The incredibly wide wavelength range, covering the Rayleigh regime as well as various potential molecular lines (CO, $CO_2$, $H_2O$), is ideally suited to accurately constraining the bulk properties of AB Pic b including the luminosity, $T_{eff}$, $\log(g)$, and radius. Both NIRSpec and MIRI will enable us to explore the potential effects of clouds and to accurately measure the molecular abundances and chemistry at play in the atmosphere. AB Pic b will serve as an ideal benchmark for comparison with young free-floating BDs, and early L-type planets like $\beta$ Pic b, HIP 65426 b (Early Release Science program, ID-1386, PI: Hinkley), and TYC 8998-760-1 b and c (Cycle 1 GO program, ID-2044, PI: Wilcomb), which share very similar ranges of mass and temperature, but are the products of potentially different formation processes.

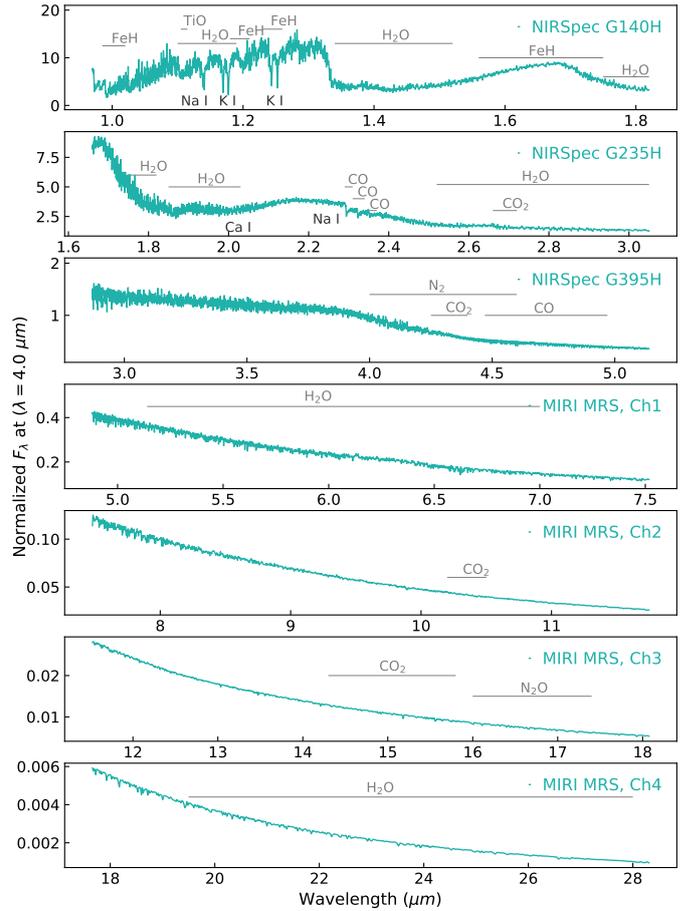

**Fig. 12.** JWST NIRSpec and MIRI predictions from Exo-REM with $T_{eff}$ = 1800 K, $\log(g)$ = 4.5 dex, [M/H] = 0.22, and $^C/_O$ = 0.6. The values adopted are not exactly the ones listed in Table 4 so as to avoid interpolations.

## 6. Summary and conclusions

To perform this study, we reduced and corrected archival $K$ band SINFONI and SPHERE H2H3-dual band data and collected available spectra and photometric points on other wavelengths. Altogether, this gives a unique set of observations covering a broad range of the SED, which is ideal for modeling the atmosphere of this companion, and there is enough time between photometric observations to allow us to partially resolve the orbit. Our definitive conclusions are as follows:

i. In general, the multi-modeling performance allowed a good derivation of the physical atmospheric properties, which we

find to be compatible with previous findings and evolutionary models, when available. We report a first estimation for the $^C/_O$ ratio of the system and a potential super-solar [M/H]. We emphasize that the Exo-REM model B7 fits the $K$ band continuum shape, lines, and especially the CO bandheads very well, making our estimation of $^C/_O = 0.58 \pm 0.08$ robust.

ii. From the orbital modeling, we find that an edge-on system configuration is favored ($i = 90$ deg), making it ideally suited for high-precision RV follow-up measurements.

iii. From the published rotation period of 2.1 h and the derived $v\sin(i)$ of $73^{+11}_{-27}$ km s$^{-1}$, we estimate the true obliquity of AB Pic b for the first time ($\Psi_{op} \in 45\text{--}135$ deg), indicating a rather significant misalignment between the planet's spin and orbit orientations.

iv. We exclude the presence of an inner massive companion to AB Pic b with a mass $\gtrsim 2\,M_{\rm Jup}$ at a physical projected separation $\gtrsim 10$ au. However, we report the existence of a significant proper-motion anomaly that could be explained by the presence of a $\sim 6\,M_{\rm Jup}$ exoplanet orbiting at less than 10 au.

Our results suggest that AB Pic b was potentially formed through gravitational instability or core accretion closer to the star before being scattered to a wide orbit by a planet–planet scattering dynamical interaction with the inner companion currently indicated by the HIPPARCOS–Gaia proper-motion anomaly of the stellar host. The confirmation of AB Pic c, along with access to broad wavelength coverage and high-resolution data for AB Pic b could position this system as an ideal benchmark for comparison when studying the formation and evolution histories of planetary systems.

*Acknowledgements.* To the memory of Dr. France Allard and her contributions to the domain of atmospheric physics of stars, brown dwarfs, and exoplanets. P.P.B. acknowledges support from the visiting program of the CNRS French-Chilean Laboratory for Astrophysics (FCLA, IRL-3386), under contract number 14177/2021. PMR acknowledges support from the ANID BASAL project ACE210002. I.S. and Y.Z. acknowledge funding from the European Research Council (ERC) under the European Union's Horizon 2020 research and innovation programme under grant agreement no. 694513. We acknowledge support in France from the French National Research Agency (ANR) through project Grant ANR-20-CE31-0012 and the Programmes Nationaux de Planétologie et de Physique Stellaire (PNP and PNPS). This project has received funding from the European Research Council (ERC) under the European Union's Horizon 2020 research and innovation programme (COBREX; Grant agreement 885 593). This publication makes use of the SIMBAD and VizieR databases operated at the CDS, Strasbourg, France. This work is partly based on data products from the SPHERE GTO processed at the SPHERE Data Centre hosted at OSUG/IPAG, Grenoble. SPHERE is an instrument designed and built by a consortium consisting of IPAG (Grenoble, France), MPIA (Heidelberg, Germany), LAM (Marseille, France), LESIA (Paris, France), Laboratoire Lagrange (Nice, France), INAF–Osservatorio di Padova (Italy), Observatoire de Genève (Switzerland), ETH Zurich (Switzerland), NOVA (Netherlands), ONERA (France) and ASTRON (Netherlands) in collaboration with ESO. SPHERE was funded by ESO, with additional contributions from CNRS (France), MPIA (Germany), INAF (Italy), FINES (Switzerland), and NOVA (Netherlands). SPHERE also received funding from the European Commission Sixth and Seventh Framework Programmes as part of the Optical Infrared Coordination Network for Astronomy (OPTICON) under Grant number RII3-Ct-2004-001566 for FP6 (2004—2008), Grant number 226604 for FP7 (2009–2012), and Grant number 312430 for FP7 (2013—2016).

# Appendix A: Orbital modeling posteriors

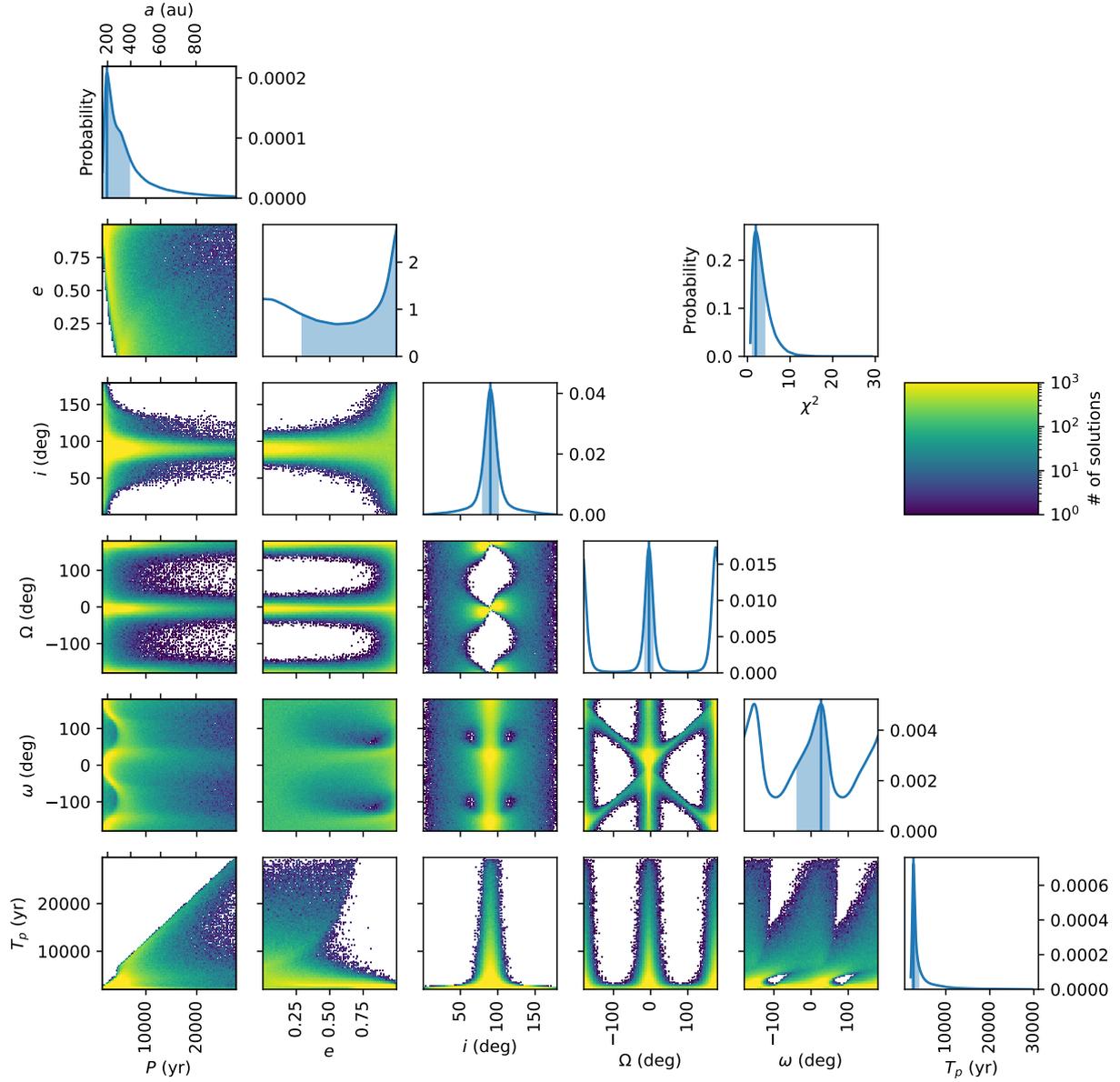

**Fig. A.1.** Corner plot showing the posteriors of the orbital fit of AB Pic b performed with OFTI, combining NaCo and SPHERE observations. The blue vertical lines in the diagonal plots indicate the peak value and the blue shaded areas indicate the shortest 68% confidence intervals.

## Appendix B: ForMoSA posteriors

In this Appendix, the two tables list the posteriors with asymmetric one-sigma confidence intervals for all the performed models. Both tables are very similar but Table B.1 is for the BT-SETTL13 family of atmospheric models while B.2 is for Exo-REM.

**Table B.1.** Multi-model approach with BT-SETTL13.

| BT-SETTL 13 | | $T_{eff}$ (K) | log(g) (dex) | $^c/_O$ | R ($R_{jup}$) | $A_v$ mag | RV ($km s^{-1}$) | v sin(i) ($km s^{-1}$) | log(z) |
|---|---|---|---|---|---|---|---|---|---|
| Priors | | (1400 − 2200) | (3.5 − 5) | (0.28 − 1.09) | (0.5 − 10) | (0 − 5) | (−10 − 100) | (0 − 500) | |
| Full SED | A0 | $1804.8^{+19.8}_{-4.4}$ | $4.00^{+0.14}_{-0.02}$ | $0.61^{+0.03}_{-0.03}$ | | | | | $-14355.7 \pm 0.2$ |
| | A1 | $1805.7^{+187.9}_{-3.2}$ | $4.0^{+0.05}_{-0.06}$ | $0.61^{+0.04}_{-0.02}$ | | $0.01^{+2.69}_{-0.01}$ | | | $-14364.7 \pm 0.2$ |
| $JHKL_p$ | A2 | $1821.5^{+13.7}_{-13.2}$ | $4.7^{+0.06}_{-0.38}$ | $0.50^{+0.09}_{-0.02}$ | | | | | $-11636.2 \pm 0.2$ |
| K band | A3 | $1657.2^{+18.8}_{-63.6}$ | $3.54^{+0.32}_{-0.08}$ | $0.74^{+0.03}_{-0.15}$ | | | | | $-1130.7 \pm 0.2$ |
| w/ cont | A4 | $1658.3^{+79.7}_{-44.9}$ | $3.61^{+0.59}_{-0.10}$ | $0.74^{+0.03}_{-0.16}$ | $1.93^{+0.12}_{-0.22}$ | | | | $-1519.9 \pm 0.2$ |
| | A5 | $1664.6^{+16.5}_{-40.8}$ | $3.55^{+0.24}_{-0.08}$ | $0.73^{+0.03}_{-0.12}$ | | | $22.3^{+23.1}_{-3.8}$ | | $-863.2 \pm 0.2$ |
| | A6 | $1670.4^{+18.0}_{-25.5}$ | $3.57^{+0.21}_{-0.08}$ | $0.72^{+0.03}_{-0.09}$ | | | $34.3^{+17.4}_{-11.1}$ | $72.0^{+127.8}_{-13.2}$ | $-782.7 \pm 0.2$ |
| | A7 | $1670.4^{+14.1}_{-20.2}$ | $3.58^{+0.27}_{-0.06}$ | $0.72^{+0.03}_{-0.07}$ | | $0.14^{+1.67}_{-0.13}$ | $34.2^{+13.4}_{-10.2}$ | $71.7^{+95.3}_{-13.1}$ | $-787.9 \pm 0.2$ |
| | A8 | $1669.2^{+17.5}_{-104.4}$ | $3.57^{+0.29}_{-0.12}$ | $0.72^{+0.03}_{-0.15}$ | $1.87^{+0.16}_{-0.07}$ | | $34.4^{+22.0}_{-13.7}$ | $74.6^{+195.9}_{-16.1}$ | $-790.9 \pm 0.2$ |
| | A9 | $1668.4^{+11.4}_{-58.4}$ | $3.59^{+0.35}_{-0.07}$ | $0.71^{+0.02}_{-0.12}$ | $1.88^{+0.24}_{-0.03}$ | $0.22^{+2.28}_{-0.21}$ | $34.5^{+19.5}_{-11.6}$ | $73.0^{+164.0}_{-13.9}$ | $-795.5 \pm 0.2$ |
| w/o cont | A10 | $1547.2^{+290.8}_{-16.5}$ | $2.76^{+0.62}_{-0.23}$ | $0.78^{+0.03}_{-0.09}$ | | | | | $-689.1 \pm 0.1$ |
| | A11 | $1734.4^{+217.0}_{-188.8}$ | $3.37^{+0.60}_{-0.66}$ | $0.68^{+0.13}_{-0.15}$ | | | $31.0^{+10.6}_{-3.4}$ | | $-520.5 \pm 0.2$ |
| | A12 | $1914.2^{+62.8}_{-266.0}$ | $3.72^{+0.63}_{-0.59}$ | $0.57^{+0.17}_{-0.08}$ | | | $33.7^{+25.5}_{-6.1}$ | $46.5^{+134.0}_{-18.5}$ | $-518.1 \pm 0.2$ |
| | A13 | $1854.5^{+113.2}_{-345.7}$ | $3.69^{+0.39}_{-0.58}$ | $0.58^{+0.22}_{-0.09}$ | $1.14^{+1.40}_{-0.14}$ | | $34.2^{+82.2}_{-7.4}$ | $55.3^{+205.9}_{-23.2}$ | $-524.2 \pm 0.2$ |
| J band | A14 | $1634.5^{+63.5}_{-8.9}$ | $4.66^{+0.1}_{-0.32}$ | $0.33^{+0.02}_{-0.04}$ | | | | | $-1104.7 \pm 0.1$ |
| w/ cont | A15 | $1633.7^{+70.5}_{-25.8}$ | $4.64^{+0.11}_{-0.51}$ | $0.36^{+0.02}_{-0.08}$ | $1.99^{+0.57}_{-0.10}$ | | | | $-1111.6 \pm 0.2$ |
| | A16 | $1635.7^{+45.4}_{-8.4}$ | $4.66^{+0.09}_{-0.18}$ | $0.32^{+0.18}_{-0.04}$ | | | $38.6^{+17.2}_{-7.5}$ | $19.9^{+34.7}_{-13.2}$ | $-1060.2 \pm 0.2$ |
| | A17 | $1634.8^{+48.5}_{-11.5}$ | $4.65^{+0.1}_{-0.37}$ | $0.34^{+0.03}_{-0.06}$ | $1.98^{+0.38}_{-0.08}$ | | $38.6^{+21.7}_{-10.8}$ | $22.0^{+41.9}_{-14.8}$ | $-1066.2 \pm 0.2$ |
| w/o cont | A18 | $1574.6^{+29.7}_{-46.3}$ | $4.60^{+0.13}_{-0.55}$ | $0.36^{+0.02}_{-0.07}$ | | | | | $-534.2 \pm 0.1$ |
| | A19 | $1570.5^{+26.5}_{-54.9}$ | $4.57^{+0.15}_{-0.63}$ | $0.38^{+0.25}_{-0.09}$ | $4.25^{+1.95}_{-0.27}$ | | | | $-538.6 \pm 0.1$ |
| | A20 | $1581.4^{+17.5}_{-42.4}$ | $4.62^{+0.11}_{-0.49}$ | $0.34^{+0.19}_{-0.05}$ | | | $38.6^{+13.0}_{-7.0}$ | $25.6^{+40.1}_{-17.1}$ | $-461.4 \pm 0.1$ |
| | A21 | $1578.2^{+20.3}_{-51.3}$ | $4.6^{+0.13}_{-0.58}$ | $0.36^{+0.02}_{-0.06}$ | $4.24^{+1.55}_{-0.23}$ | | $38.7^{+14.9}_{-8.1}$ | $28.6^{+41.4}_{-19.6}$ | $-465.2 \pm 0.2$ |
| H band | A22 | $1895.6^{+16.4}_{-171.6}$ | $4.01^{+0.85}_{-0.22}$ | $0.97^{+0.02}_{-0.28}$ | | | | | $-210.5 \pm 0.1$ |
| | A23 | $1878.8^{+32.2}_{-274.0}$ | $4.06^{+0.78}_{-0.27}$ | $0.93^{+0.06}_{-0.35}$ | $1.34^{+0.81}_{-0.07}$ | | | | $-217.3 \pm 0.2$ |
| $L_p$ band | A24 | $1950.3^{+21.1}_{-355.0}$ | $3.57^{+0.32}_{-0.07}$ | $0.92^{+0.03}_{-0.17}$ | | | | | $-1070.5 \pm 0.1$ |
| | A25 | $1872.5^{+100.3}_{-281.7}$ | $3.63^{+0.62}_{-0.12}$ | $0.87^{+0.08}_{-0.21}$ | $1.29^{+0.28}_{-0.03}$ | | | | $-1078.1 \pm 0.2$ |
| Adopted Values | | $1800 \pm 20$ | $4.6^{+0.1}_{-0.5}$ | $0.71 \pm 0.15$ | $1.8 \pm 0.2$ | - | $33 \pm 10$ | $75^{+200}_{-16}$ | |

**Notes.** The priors are listed in the first row. We explored the outcomes by varying wavelength ranges and the number of free parameters. The models are labeled with an A from 0 to 25 and the adopted values from this work and this atmospheric grid are listed in the last row. The natural logarithm of the Bayesian evidence for each model is listed in the last column of each row, together with its estimated numerical (sampling) error. A0, A8, and A20 are the models presented in the text and we colored them to match the colors in Figures 6 and 7.

**Table B.2.** Multi-model approach with Exo-REM.

| Exo-REM | $T_{eff}$ (K) | log(g) (dex) | [M/H] | $C/O$ | R ($R_{jup}$) | $A_v$ mag | RV ($kms^{-1}$) | v sin(i) ($kms^{-1}$) | log(z) |
|---|---|---|---|---|---|---|---|---|---|
| Priors | (400 – 2000) | (3 – 5) | (−0.5 – 1) | (0.1 – 0.8) | (0.5 – 10) | (0 – 5) | (−10 – 100) | (0 – 500) | |
| Full SED B0 | $1700.0^{+1.7}_{-55.8}$ | $3.50^{+0.01}_{-0.08}$ | $1.00^{+0.01}_{-0.26}$ | $0.67^{+0.01}_{-0.32}$ | | | | | $-8969.6 \pm 0.3$ |
| B1 | $1989.7^{+5.7}_{-133.6}$ | $3.50^{+0.07}_{-0.09}$ | $0.18^{+0.54}_{-0.01}$ | $0.70^{+0.01}_{-0.33}$ | | $3.25^{+0.07}_{-1.06}$ | | | $-7455.5 \pm 0.2$ |
| $JHKL_p$ B2 | $1698.9^{+15.5}_{-69.3}$ | $3.53^{+0.11}_{-0.23}$ | $0.99^{+0.02}_{-0.40}$ | $0.78^{+0.02}_{-0.22}$ | | | | | $-7295.2 \pm 0.2$ |
| K band B3 | $1776.1^{+9.7}_{-313.2}$ | $3.31^{+0.87}_{-0.03}$ | $0.77^{+0.92}_{-0.04}$ | $0.84^{+0.02}_{-0.22}$ | | | | | $-1086.4 \pm 0.2$ |
| w/ cont B4 | $1774.5^{+7.0}_{-226.4}$ | $3.27^{+0.76}_{-0.04}$ | $0.77^{+0.31}_{-0.12}$ | $0.81^{+0.03}_{-0.25}$ | | | $19.0^{+4.4}_{-4.0}$ | | $-728.5 \pm 0.2$ |
| B5 | $1798.7^{+85.0}_{-168.3}$ | $4.03^{+0.23}_{-0.41}$ | $0.52^{+0.57}_{-0.18}$ | $0.58^{+0.09}_{-0.06}$ | | | $27.0^{+11.9}_{-10.9}$ | $74.5^{+10.4}_{-16.5}$ | $-560.4 \pm 0.2$ |
| B6 | $1892.0^{+96.4}_{-254.2}$ | $3.82^{+0.47}_{-0.21}$ | $0.53^{+0.54}_{-0.18}$ | $0.64^{+0.03}_{-0.12}$ | | $1.17^{+0.69}_{-2.69}$ | $26.9^{+11.6}_{-9.4}$ | $74.3^{+10.0}_{-16.0}$ | $-558.9 \pm 0.2$ |
| B7 | $1758.0^{+47.6}_{-320.9}$ | $3.96^{+0.16}_{-0.49}$ | $0.36^{+0.21}_{-0.16}$ | $0.58^{+0.08}_{-0.08}$ | $1.70^{+0.76}_{-0.06}$ | | $28.5^{+21.9}_{-8.8}$ | $73.1^{+11.1}_{-26.9}$ | $-560.7 \pm 0.5$ |
| B8 | $1805.9^{+180.3}_{-285.4}$ | $3.76^{+0.37}_{-0.37}$ | $0.33^{+0.23}_{-0.14}$ | $0.63^{+0.04}_{-0.13}$ | $1.64^{+0.64}_{-0.16}$ | $1.06^{+1.26}_{-2.43}$ | $28.3^{+18.5}_{-7.3}$ | $73.7^{+10.8}_{-23.9}$ | $-560.4 \pm 0.5$ |
| w/o cont B9 | $1922.7^{+75.6}_{-172.7}$ | $4.67^{+0.25}_{-0.59}$ | $0.59^{+0.39}_{-0.19}$ | $0.65^{+0.15}_{-0.15}$ | | | | | $-743.5 \pm 0.1$ |
| B10 | $1936.4^{+58.2}_{-183.1}$ | $4.81^{+0.17}_{-0.82}$ | $0.46^{+0.15}_{-0.28}$ | $0.50^{+0.08}_{-0.05}$ | | | $30.5^{+3.4}_{-2.3}$ | | $-717.1 \pm 0.2$ |
| B11 | $1950.1^{+46.0}_{-160.1}$ | $4.79^{+0.18}_{-0.72}$ | $0.48^{+0.13}_{-0.34}$ | $0.50^{+0.05}_{-0.05}$ | | | $35.6^{+4.0}_{-3.7}$ | $43.8^{+18.8}_{-2.5}$ | $-555.3 \pm 0.2$ |
| B12 | $1933.0^{+62.2}_{-282.2}$ | $4.70^{+0.26}_{-0.61}$ | $0.47^{+0.15}_{-0.38}$ | $0.50^{+0.07}_{-0.08}$ | $1.03^{+0.46}_{-0.09}$ | | $35.7^{+5.5}_{-4.3}$ | $43.9^{+23.1}_{-3.5}$ | $-560.2 \pm 0.2$ |
| J band B13 | $1606.9^{+124.3}_{-33.3}$ | $4.71^{+0.14}_{-0.94}$ | $0.50^{+0.22}_{-0.24}$ | $0.20^{+0.26}_{-0.01}$ | | | | | $-1045.0 \pm 0.2$ |
| w/ cont B14 | $1602.0^{+74.5}_{-296.7}$ | $4.68^{+0.16}_{-0.95}$ | $0.48^{+0.23}_{-0.40}$ | $0.22^{+0.31}_{-0.27}$ | $1.83^{+1.73}_{-0.27}$ | | | | $-1051.6 \pm 0.2$ |
| B15 | $1608.3^{+141.4}_{-249.6}$ | $4.73^{+0.14}_{-0.66}$ | $0.51^{+0.22}_{-0.24}$ | $0.20^{+0.21}_{-0.02}$ | | | $37.2^{+11.7}_{-8.9}$ | $21.4^{+33.0}_{-14.1}$ | $-1009.7 \pm 0.5$ |
| B16 | $1508.4^{+92.8}_{-249.6}$ | $4.71^{+0.15}_{-0.66}$ | $0.44^{+0.24}_{-0.45}$ | $0.26^{+0.31}_{-0.07}$ | $2.19^{+1.92}_{-0.35}$ | | $37.6^{+17.8}_{-11.4}$ | $27.5^{+32.4}_{-16.3}$ | $-1019.0 \pm 0.7$ |
| w/o cont B17 | $1179.2^{+79.2}_{-28.2}$ | $4.75^{+0.22}_{-0.58}$ | $-0.38^{+0.04}_{-0.11}$ | $0.74^{+0.03}_{-0.22}$ | | | | | $-498.8 \pm 0.1$ |
| B18 | $1193.9^{+85.1}_{-39.4}$ | $4.79^{+0.17}_{-0.66}$ | $-0.28^{+0.68}_{-0.09}$ | $0.74^{+0.03}_{-0.28}$ | $9.27^{+0.58}_{-2.98}$ | | | | $-504.6 \pm 0.2$ |
| B19 | $1217.1^{+47.4}_{-52.9}$ | $4.51^{+0.23}_{-0.30}$ | $-0.36^{+0.23}_{-0.13}$ | $0.67^{+0.08}_{-0.11}$ | | | $35.6^{+9.1}_{-8.2}$ | $22.8^{+38.7}_{-15.0}$ | $-446.3 \pm 0.2$ |
| B20 | $1197.8^{+103.5}_{-36.1}$ | $4.76^{+0.10}_{-0.61}$ | $-0.27^{+0.75}_{-0.10}$ | $0.74^{+0.04}_{-0.26}$ | $9.34^{+0.54}_{-2.83}$ | | $35.5^{+10.6}_{-9.6}$ | $22.1^{+41.4}_{-14.5}$ | $-456.8 \pm 0.2$ |
| H band B21 | $1919.2^{+60.0}_{-411.5}$ | $3.83^{+0.57}_{-0.41}$ | $0.31^{+0.53}_{-0.40}$ | $0.48^{+0.23}_{-0.21}$ | | | | | $-190.6 \pm 0.1$ |
| B22 | $1857.7^{+122.9}_{-598.8}$ | $3.87^{+0.65}_{-0.45}$ | $0.24^{+0.55}_{-0.41}$ | $0.51^{+0.20}_{-0.24}$ | $1.38^{+2.02}_{-0.14}$ | | | | $-197.7 \pm 0.1$ |
| $L_p$ band B23 | $1692.1^{+119.9}_{-60.4}$ | $3.54^{+0.37}_{-0.31}$ | $0.78^{+0.15}_{-0.62}$ | $0.77^{+0.02}_{-0.46}$ | | | | | $-1017.0 \pm 0.2$ |
| B24 | $1686.4^{+111.4}_{-256.5}$ | $3.56^{+0.63}_{-0.31}$ | $0.72^{+0.20}_{-0.69}$ | $0.70^{+0.08}_{-0.42}$ | $1.64^{+0.36}_{-0.14}$ | | | | $-1024.1 \pm 0.2$ |
| Adopted Values | $1700 \pm 50$ | $4.5 \pm 0.3$ | $0.36 \pm 0.20$ | $0.58 \pm 0.08$ | $1.7^{+0.8}_{-0.1}$ | - | $32 \pm 6$ | $73^{+11}_{-27}$ | |

**Notes.** This table is analogous to Table B.1 but for Exo-REM where [M/H] variations are explored by the models. The labels are B from 0 to 24. B0, B7 and, B19 are the best outcomes represented in Figures 6 and 8.